\documentclass[prb,amsmath,amssymb,amsbsy,twocolumn,superscriptaddress,showpacs]{revtex4-1}

\usepackage{amsmath,amssymb}
\usepackage{wasysym}
\usepackage{graphics}
\usepackage[pdftex]{graphicx}
\usepackage{epstopdf}
\usepackage{latexsym}
\usepackage{subfigure}
\usepackage{color}
\usepackage{multirow}
\usepackage[]{natbib}
\usepackage[breaklinks,colorlinks = true,linkcolor = red,urlcolor=cyan,citecolor=red]{hyperref}
\usepackage{times}
\usepackage[abs]{overpic}

\newcommand{\be}{\begin{equation}}
\newcommand{\ee}{\end{equation}}
\newcommand{\bw}{\begin{widetext}}
\newcommand{\ew}{\end{widetext}}
\newcommand{\bea}{\begin{eqnarray}}
\newcommand{\eea}{\end{eqnarray}}
\newcommand{\ba}{\begin{array}}
\newcommand{\ea}{\end{array}}

\newcommand{\nn}{\nonumber}
\newcommand{\bs}{\boldsymbol}
\newcommand{\la}{\langle}
\newcommand{\ra}{\rangle}

\newcommand{\sslash}{\mathbin{/\mkern-6mu/}}

\begin{document}

\title{Emergent quantum phases in a frustrated $\bs{J_1}$-$\bs{J_2}$ Heisenberg model on 
hyperhoneycomb lattice}

\author{SungBin Lee}
\affiliation{Department of Physics, University of Toronto, Toronto, Ontario M5S 1A7, Canada}
\author{Jae-Seung Jeong}
\affiliation{School of Physics, Korea Institute for Advanced Study, Seoul 130-722, Korea}
\author{Kyusung Hwang}
\affiliation{Department of Physics, University of Toronto, Toronto, Ontario M5S 1A7, Canada}
\author{Yong Baek Kim}
\affiliation{Department of Physics, University of Toronto, Toronto, Ontario M5S 1A7, Canada}
\affiliation{School of Physics, Korea Institute for Advanced Study, Seoul 130-722, Korea}

\date{\today}
\begin{abstract}
We investigate possible quantum ground states as well as the classical limit of a frustrated 
$J_1$-$J_2$ Heisenberg model on the three-dimensional (3D) hyperhoneycomb lattice. 
Our study is inspired by the recent discovery of $\beta$-Li$_2$IrO$_3$,\cite{2013_takagi} 
where Ir$^{4+}$ ions form a 3D network with each lattice site being connected to three nearest neighbors.
We focus on the influence of magnetic frustration caused by the second-nearest neighbor 
spin interactions. Such interactions are likely to be significant 
due to large extent of 5d orbitals in iridates or other 5d transition metal oxides. 
In the classical limit, the ground state manifold is given by 
line degeneracies of the spiral magnetic-order wavevectors when $ J_2/J_1\gtrsim 0.17$ 
while the collinear stripy order is included in the degenerate manifold when $J_2/J_1 = 0.5$. 
Quantum order-by-disorder effects are studied using both the semi-classical $1/S$ 
expansion in the spin wave theory and Schwinger boson approach. 
In general, certain coplanar spiral orders are chosen from the classical degenerate manifold for
a large fraction of the phase diagram.
Nonetheless quantum fluctuations favor the collinear stripy order over the spiral orders 
in an extended parameter region around $J_2/J_1 = 0.5$, despite the spin-rotation invariance
of the underlying Hamiltonian. This is in contrast to the emergence of stripy order in the 
Heisenberg-Kitaev model studied earlier on the same lattice, where
the Kitaev-type Ising interactions are important for stabilizing the stripy order.
\cite{PhysRevB.89.045117,2013arXiv1309.1171K,PhysRevB.89.014424} 
As quantum fluctuations become stronger, U(1) and Z$_2$ quantum spin 
liquid phases are shown to arise via quantum disordering of the N\'eel, stripy and 
spiral magnetically-ordered phases. The effects of magnetic anisotropy and their relevance 
to future experiments are also discussed.
\end{abstract}
\maketitle

\section{introduction}
\label{sec:intro}

Recent research activities on 5d transition metal oxides suggest that the strong spin-orbit
coupling in conjunction with electron correlation may lead to unusual topological and magnetic
phases.\cite{PhysRevLett.102.017205, PhysRevB.82.064412, pesin2010mott, PhysRevB.82.085111, PhysRevB.85.045124,PhysRevLett.105.027204,
PhysRevB.83.220403, PhysRevLett.108.127203, witczak2013correlated, PhysRevLett.110.076402}  
In particular, it has been recognized that the band-width of spin-orbit-reorganized
bands near the Femi level often becomes relatively narrow and moderate strength of
electron correlation may be enough to generate Mott insulators.
In such 5d Mott insulators, however, one would expect that spin exchange 
interactions between lattice sites beyond nearest-neighbors would become significant 
due to the extended nature of 5d orbitals. In addition, many of these Mott insulators
are the so-called weak Mott insulators with a small charge gap and 
significant local charge fluctuations can generate multi-spin exchange interactions
around multiple lattice sites.\cite{PhysRevLett.109.266406, dai2013local,
:/content/aip/journal/apl/96/18/10.1063/1.3374449,PhysRevLett.108.127204,PhysRevLett.110.076402} 
These further neighbor spin interactions, for example,
may provide magnetic frustration and lead to emergence of quantum spin liquid 
and/or other exotic phases. 
On the other hand, various forms of anisotropic spin interactions may also be present
due to spin-orbit coupling. The interplay between anisotropic spin interactions 
and the magnetic frustration effect is of fundamental importance in 
understanding quantum magnetism in 5d transition metal oxides.

The layered two-dimensional (2D) honeycomb iridates, Na$_2$IrO$_3$ and Li$_2$IrO$_3$, are fertile 
playgrounds, where the combined effects of anisotropic spin interactions and
extended nature of 5d orbitals have been intensively studied.
\cite{PhysRevB.82.064412, PhysRevB.83.220403, PhysRevLett.108.127203,
PhysRevLett.105.027204, PhysRevLett.110.097204}  
For example, it is shown that the so-called Kitaev model with bond-dependent anisotropic spin 
interactions may arise in the strong Mott regime. This has raised the hope that
a quantum spin liquid phase, the exact solution of the model, may be realized.
\cite{kitaev2006anyons,2013arXiv1303.5922N} 
However, the ground state phase diagram of the Kitaev-Heisenberg 
model, where the Kitaev term is supplemented by the antiferromagnetic nearest-neighbor 
Heisenberg interaction, is not fully consistent with the magnetic orders discovered in
Na$_2$IrO$_3$ and Li$_2$IrO$_3$. It has been suggested that further neighbor
spin interactions due to the large extent of 5d orbitals may also be important for
the explanation of the experimental results.\cite{PhysRevB.84.180407, PhysRevB.88.035107} 
Which interaction would play the dominant
role for the determination of magnetic order or ground state in real materials 
has been a subject of intensive debate.

In this paper, we study a frustrated $J_1$-$J_2$ Heisenberg model on the three-dimensional (3D) hyperhoneycomb
lattice, where Ir$^{4+}$ ions reside in the newly discovered 3D iridate $\beta$-Li$_2$IrO$_3$.\cite{2013_takagi}
Here $J_1$ and $J_2$ represent the nearest- and next-nearest neighbor exchange interactions.
In the hyperhoneycomb lattice, each lattice site is connected to three neighboring sites just  
like the 2D honeycomb lattice. The ground state phase diagram of the Heisenberg-Kitaev model 
on this lattice has recently been studied and contains a number of collinear magnetic orders
as well as a 3D quantum-spin-liquid in the Kitaev limit.\cite{PhysRevB.79.024426, PhysRevB.89.045117,
PhysRevB.89.014424,2013arXiv1309.1171K,2013arXiv1309.3068N,2014arXiv1401.7678H} 
Similarly to the 2D cousin, the full understanding of the magnetic phase diagram for $\beta$-Li$_2$IrO$_3$ 
would require the consideration of both Kitaev-like magnetic anisotropy and magnetic frustration
due to further neighbor exchange interactions.
In this work, we mostly focus on the effect of the latter by studying the simplest frustrated
spin model, where we consider the spin interactions between second nearest-neighbor sites that are
connected to a common nearest neighbor site. As we show below, this minimal model
is microscopically motivated and exhibits degenerate classical ground state manifold, a hallmark
of frustrated magnets.   

We first investigate the classical limit of this model and identify the degenerate classical 
ground state manifold. Then quantum order-by-disorder effects are studied using semi-classical 
spin-wave analysis via the $1/S$ expansion and Schwinger boson
mean-field theory. In the classical limit, the Luttinger-Tisza and single-{\bf Q} variational 
ansatz reveal various line degeneracies of ordering wave vectors for spiral magnetic order when 
$  J_2  \gtrsim 0.17 J_1$. On the other hand, 
the N\'eel order is chosen for $J_2 \lesssim 0.17 J_1$.
Interestingly, the collinear stripy order is in the classical degenerate ground state manifold at a single point $J_2=0.5 J_1$. 
This is due to the peculiar lattice geometry of the hyperhoneycomb
lattice as discussed later.

Upon including zero-point quantum fluctuations via the $1/S$ expansion in 
the spin-wave analysis, quantum order by disorder effects lift the line degeneracies in the classical spiral 
order regimes and in general select certain coplanar spiral order.
Surprisingly, the collinear stripy order wins over the spiral magnetic order for an extended 
region around $J_2/J_1=0.5$, not just at $J_2/J_1 = 0.5$. It is remarkable
that quantum fluctuations favor a collinear stripy ordered state even though the
underlying Hamiltonian is SU(2) symmetric. This is in contrast to the emergence
of the stripy order discovered earlier in the Heisenberg-Kitaev model on the same
3D lattice, where the anisotropic Ising-type spin interaction is important for the 
stabilization of the stripy order.\cite{PhysRevB.79.024426, PhysRevB.89.045117,
PhysRevB.89.014424, 2013arXiv1309.1171K} 
This suggests that {\sl if the stripy order were observed
in experiments, it could have arisen from two completely different kinds of interactions}.
The Schwinger boson analysis in the semi-classical limit corroborates the results of 
the $1/S$ expansion and provides the same general trend of the quantum order-by-disorder effect 
while the phase boundaries between different phases are not the same.
When quantum fluctuations become stronger, the Schwinger boson mean-field
theory predicts the existence of the U(1) and Z$_2$ quantum spin liquid phases
that can be obtained by quantum disordering the N\'eel, stripy and spiral magnetic ordered
phases, respectively. Finally the effects of possible magnetic anisotropy on this model 
are investigated.

The rest of the paper is organized as follows.
In Sec.~\ref{sec:J1-J2-heisenberg}, 
we introduce the $J_1$-$J_2$ Heisenberg model
on the hyperhoneycomb lattice and discuss its symmetry properties.
Using the Luttinger-Tisza and single-${\bf Q}$ variational methods, 
we determine the degenerate classical ground state manifold in Sec.~\ref{sec:classical}. 
In Sec.~\ref{sec:semi-classical}, we investigate quantum order-by-disorder effects 
on the degenerate spiral states by computing zero-point quantum fluctuations 
via $1/S$ expansion and solving Schwinger boson mean-field theory.
We show the emergence of stripy order due to quantum fluctuations.
Moreover, possible spin liquid phases in the presence of strong quantum fluctuations 
are examined using the Schwinger boson analysis.
We conclude in Sec.~\ref{sec:discussion} with a summary of our results and 
discussion on magnetic anisotropy effect.

\section{$J_1$-$J_2$ Heisenberg model on the hyperhoneycomb lattice}
\label{sec:J1-J2-heisenberg}

\begin{figure}[t!]
\centerline{\includegraphics[scale=0.5]{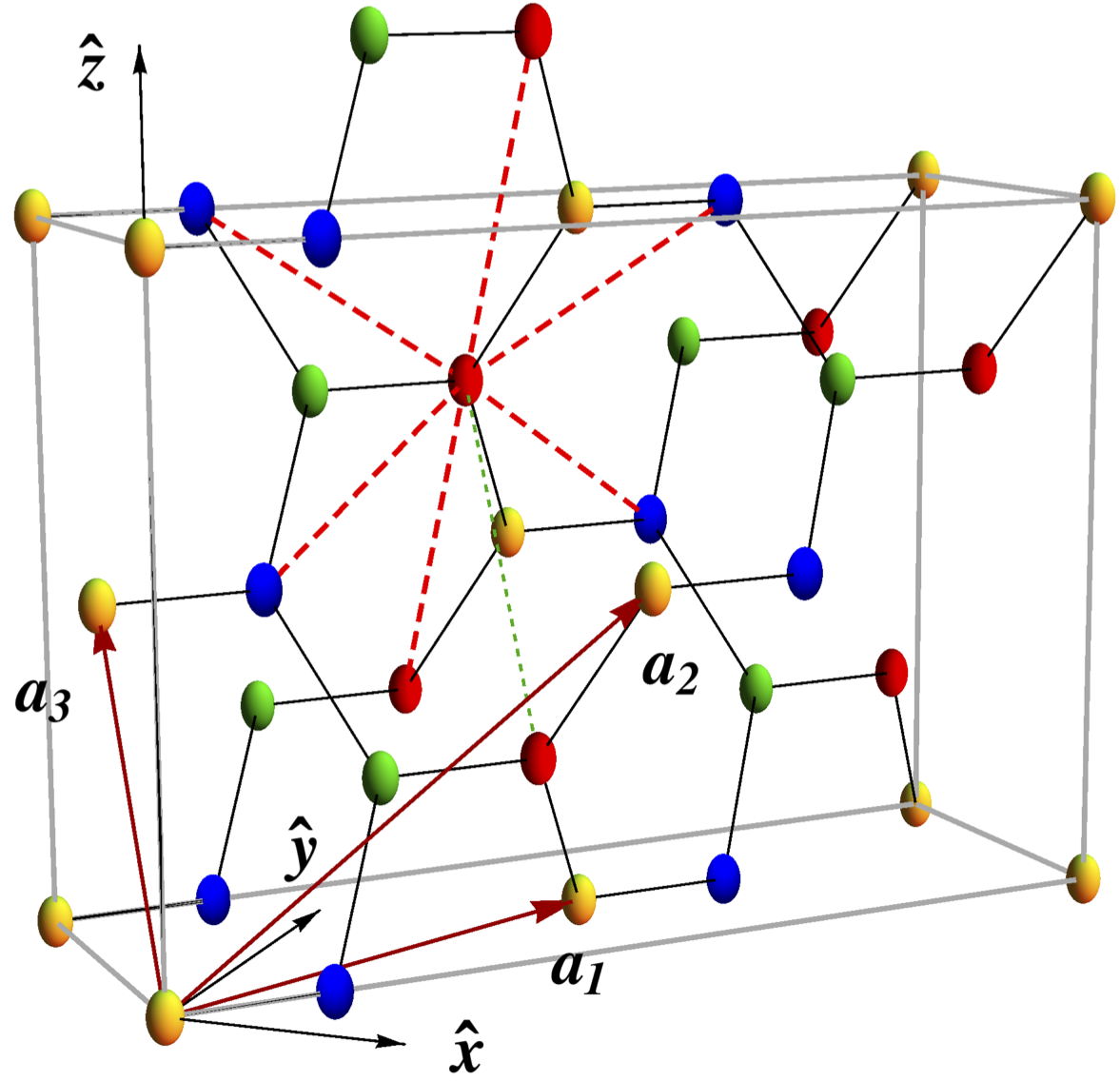}}
\caption{(Color online) The hyperhoneycomb lattice structure with tri-coordinated
four sublattices (yellow, blue, green, and red spheres).
$\bs{a}_i$ ($i=1, 2, 3$, red arrows) denote the primitive lattice vectors for 
the face-centered orthorhombic Bravais lattice.
Black solid lines show the nearest-neighbor bonds and 
red dashed lines indicate the next-nearest-neighbors that are connected 
via two nearest-neighbor bonds. 
Green dotted line has the same length as the distance between 
the next-nearest-neighbors connected by red dashed lines, but 
two sites coupled by green dotted lines are not connected 
via two nearest-neighbor bonds.}
\label{fig:hyperhoneycomb}
\end{figure}

We start by introducing the $J_1$-$J_2$ Heisenberg spin model on the hyperhoneycomb 
lattice. The model Hamiltonian is written as 
\bea
\mathcal{H} = J_1 \sum_{\la i j  \ra} {\bf S}_i \cdot {\bf S}_j + J_2 \sum_{\la \la  i j \ra \ra} {\bf S}_i \cdot {\bf S}_j ,
\label{eq:1}
\eea
where $\la i j \ra$ and $\la \la i j \ra \ra$ run over the nearest-neighbor and next-nearest-neighbor bonds, respectively.
Figure \ref{fig:hyperhoneycomb} shows the hyperhoneycomb lattice structure with four sublattices. 
Different sublattice sites labeled by $s=0,1,2,3$ are colored in yellow, blue, green and red.
This three-dimensional lattice can be regarded as a face-centered orthorhombic Bravais lattice with a four-site basis
(See Appendix~\ref{app-hyperhoneycomb} for details). Notice that each site is connected to three nearest-neighbor
sites, just like the 2D honeycomb lattice.

The nearest-neighbor bonds connect two kinds of sublattices, namely 
even ($s=0,2$) and odd sublattices ($s=1,3$). 
On the other hand, there are six next-nearest-neighbors connected 
via two nearest-neighbor bonds through a common nearest-neighbor site [red dashed lines in Fig.~\ref{fig:hyperhoneycomb}].
The connection amongst the next-nearest-neighbors exists only between even and even, or odd and odd sublattices.
Such special connectivity leads to the $J_1$-$J_2$ Heisenberg Hamiltonian ${\cal H}$ in Eq.\eqref{eq:1}
to be invariant under the following transformation
\bea
{\bf S}_{i}^{~{0, 2 (1, 3)}} \rightarrow - {\bf S}_{i}^{~{0, 2 (1, 3)}}, ~~ J_1 \rightarrow - J_1 \ ,
\label{eq:model-2}
\eea
where ${\bf S}_i^{\alpha}$ corresponds to ${\bf S}_i$ located on the sublattice $\alpha$ 
at site $i$ spanned by the primitive lattice vectors.     
Notice that the sign change of the nearest-neighbor exchange coupling $J_1$ 
is equivalent to that of the spin on either even or odd sublattices. 
Hence, without losing generality, one can explore
either ferromagnetic or antiferromagnetic sides of the magnetic phase diagram for $J_1$ 
({\it i.e.} $J_1 <0$ or $J_1 >0$), 
then the other side of the phase diagram is automatically determined, followed by 
the transformation in Eq.~\eqref{eq:model-2}. 
Throughout this paper, we assume $J_1,$ $J_2>0$ unless specified otherwise.
In addition to six next-nearest-neighbors connected by two nearest-neighbor bonds,
there are other four next-nearest-neighbors with the same lattice distance [green dotted line in Fig.~\ref{fig:hyperhoneycomb}],
but not being connected by two nearest-neighbor bonds.
Previous microscopic consideration of the underlying tight-binding model suggests that
the spin exchange interactions for the six next-nearest-neighbors are dominant and
those for the four extra next-nearest-neighbors is negligible for edge-sharing oxygen
octahedra environment of Ir$^{4+}$ ions.\cite{lee2014topological}
Hence, we focus on six next-nearest-neighbors that are connected by 
two nearest-neighbor bonds. 

\section{Magnetic phase diagram in the classical limit}
\label{sec:classical}

We first explore the classical ground states of the antiferromagnetic 
$J_1$-$J_2$ Heisenberg model [Eq.~\eqref{eq:1}] on the hyperhoneycomb lattice. 
Using two different approaches: 
Luttinger-Tisza and single-${\bf Q}$ variational analyses, 
we determine the degenerate classical-ground-state manifold.
When $J_2 \gtrsim 0.17 J_1$, we find line degeneracies of the wavevectors for 
spiral ordered phases while the N{\'e}el order is the ground state for $J_2 < 0.17 J_1$.
Notably at a single point $J_2 = 0.5 J_1$, the collinear stripy order coexists with spiral orders in the 
degenerate ground state manifold, which is shown to arise from the peculiar lattice geometry of 
the hyperhoneycomb lattice. We explain below the results of the Luttinger-Tisza approach 
and single-${\bf Q}$ variational ansatz.

\subsection{Luttinger-Tisza method}
\label{subsec:lutt-tisza}

In the nearest-neighbor antiferromagnetic Heisenberg model (for $J_1 > 0$, $J_2=0$), 
the N\'eel order is the unique ground state, where spins in even and odd sublattices 
are pointing opposite directions. 
However, when $J_2$ becomes finite and comparable to the 
strength of $J_1$, two exchange interactions compete with each other and 
may induce non-trivial magnetic order.
In this section, we adopt the Luttinger-Tisza analysis and 
investigate frustration-induced magnetic phases in the classical limit. 
The Luttinger-Tisza method~\cite{PhysRev.126.540,PhysRev.70.954,PhysRev.81.1015} 
finds the ordering wavevectors of classical spin ground states 
by minimizing the energy of ${\cal H}$ [Eq.\eqref{eq:1}],
where the spin is regarded as a classical three-component unit-vector.
The solutions are typically found in a mean-field fashion in the sense that the 
hard spin constraint, $| {\bf S}_i |=1$ at every site $i$, is only satisfied on average or a soft spin constraint, 
$\sum_{i \in uc} |{\bf S}_i|^2=N_s$ ($uc$ = unit cell and $N_s=4$ is the number of sites in the unit cells) is used.
While the solutions include the true ground state manifold,
some of the solutions may fail to satisfy the hard spin constraint.
Hence, we first look for the Luttinger-Tisza solutions with the soft spin constraint
and then find the true ground state manifold among them by examining the 
hard spin constraint.

\begin{figure}[t!]
\centerline{\includegraphics[width=4.9cm]{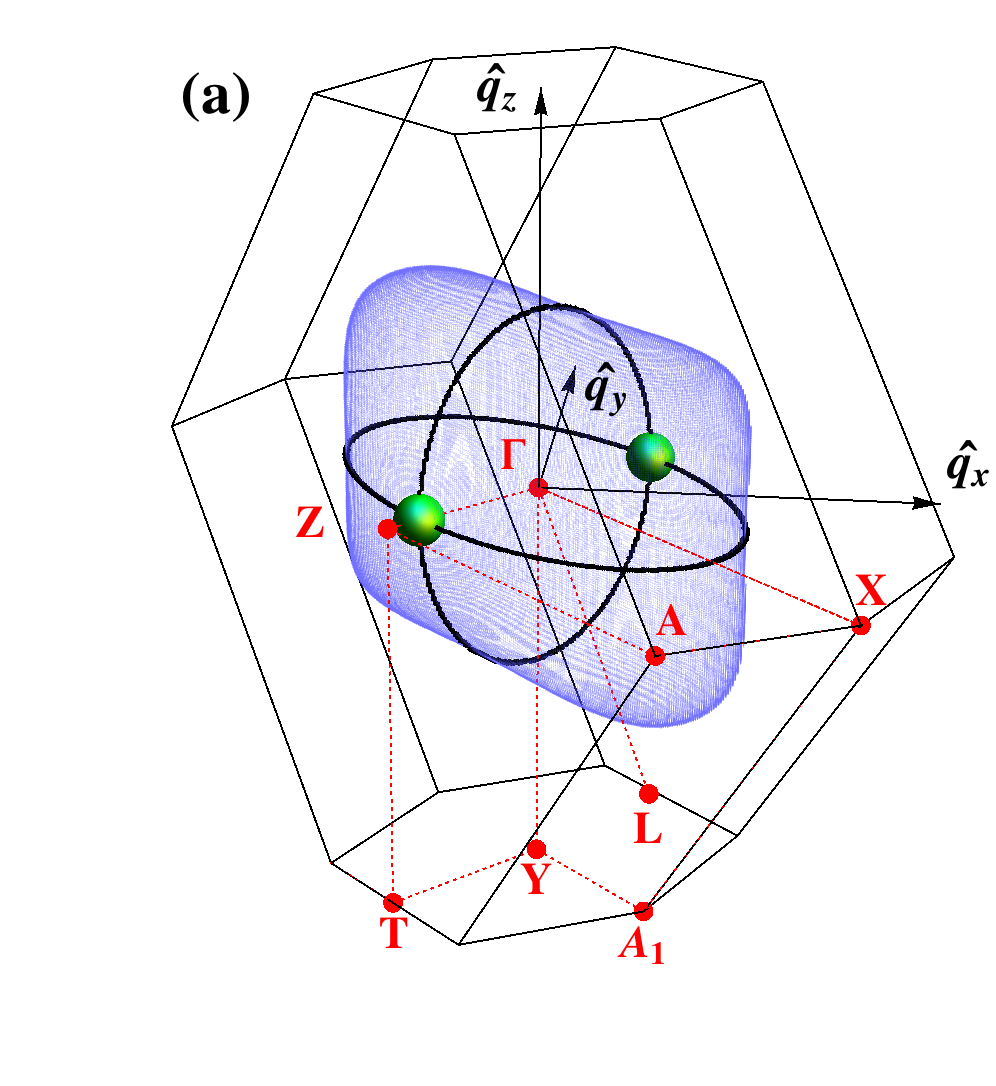}\includegraphics[width=4.9cm]{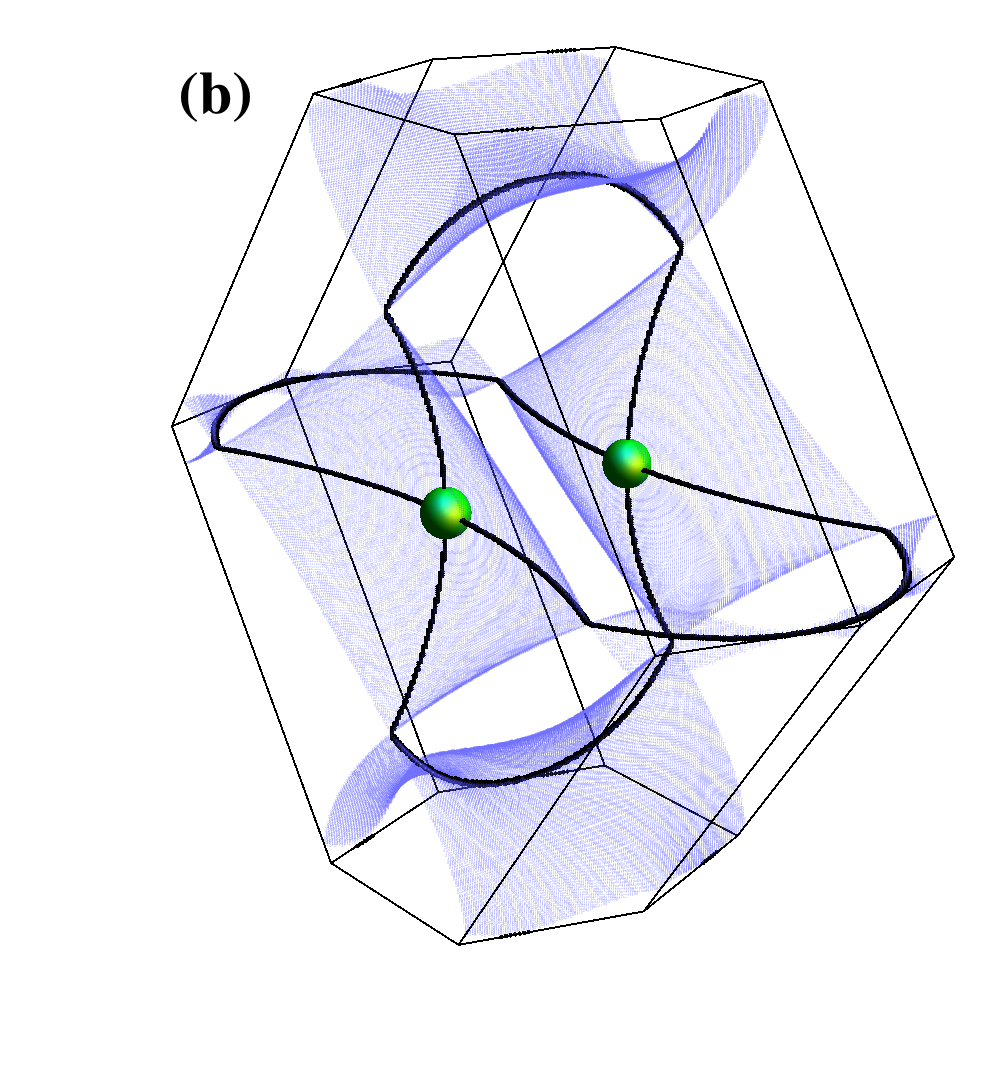}}
\centerline{\includegraphics[width=4.9cm]{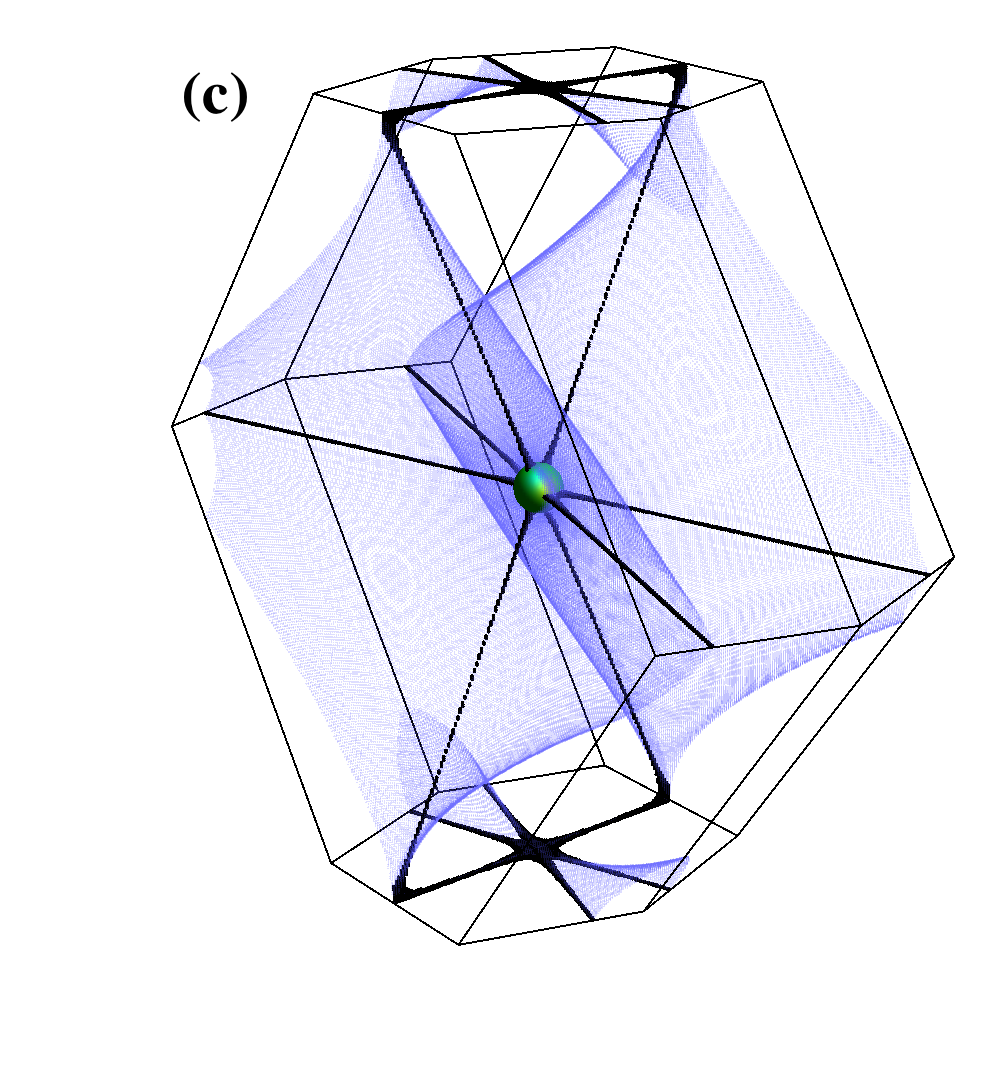}\includegraphics[width=4.9cm]{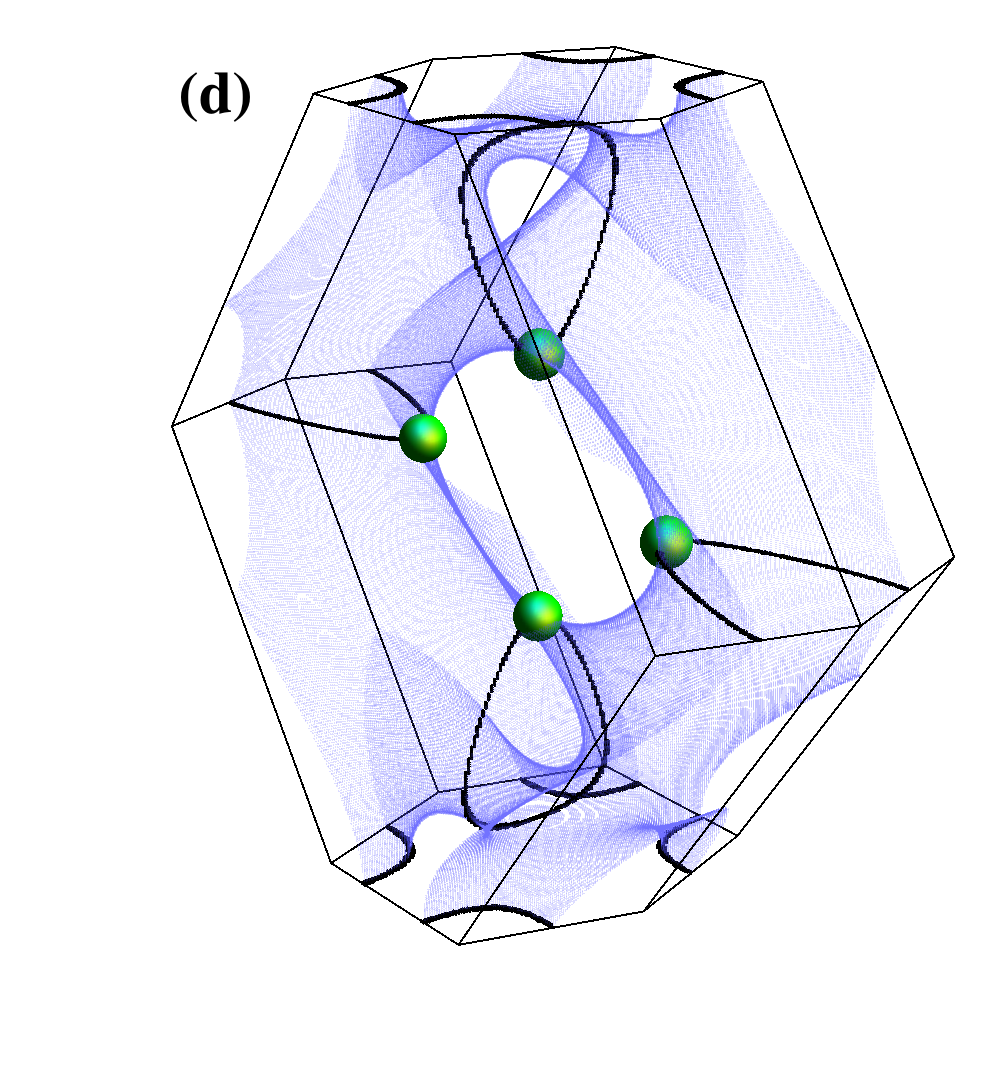}}
\caption{(Color online) Ordering wavevectors of the classical solutions obtained in the Luttinger-Tisza analysis with the soft spin constraint
are shown as degenerate surfaces (blue) for (a) $J_2/J_1=0.2$, (b) $J_2/J_1=0.3$, (c) $J_2/J_1=0.5$, and (d) $J_2/J_1=0.7$.
Black lines on the degenerate surfaces denote the ordering wavevectors of the classical spin states that satisfy the hard spin constraint.
These degenerate lines of wavevectors are consistent with the results of the single-${\bf Q}$ variational method [see Sec.~\ref{subsec:single-q}]. Green spheres represent the selected ordering wavevectors in the presence of zero-point quantum fluctuations. 
[see Sec.~\ref{subsec:HP-bosons} for details]}
\label{fig:J1_J2_LT_wavevector[1]}
\end{figure}

Based on the Luttinger-Tisza analysis, we find
that the N\'eel order with the ordering wavevector at the $\Gamma$ point, 
is the ground state for $J_2/J_1 \lesssim 0.17$. 
For $J_2/J_1\gtrsim 0.17$, on the other hand, the competition between $J_1$ and $J_2$ 
leads to degenerate spiral states, 
with the ordering wavevectors located away from the $\Gamma$ point.
Within the soft spin constraint, the degenerate ordering wavevectors, that minimize the energy of 
${\cal H}$ [Eq.\eqref{eq:1}], form a surface in the wavevector space. 
Figure \ref{fig:J1_J2_LT_wavevector[1]} shows such degenerate surfaces (colored in blue) 
of the ordering wavevectors in the first Brillouin zone.
For different values of $J_2/J_1 =0.2,0.3,0.5$ and $0.7$, 
the degenerate surfaces are shown in each panel of Figs.\ref{fig:J1_J2_LT_wavevector[1]} (a)-(d), respectively. 
We find degenerate surfaces with different topologies depending on the ratio of $J_2/J_1$. 
The degenerate wavevectors form closed surfaces in the range $0.17 \lesssim J_2/J_1 \lesssim 0.2$ and one of them is shown
in Fig.~\ref{fig:J1_J2_LT_wavevector[1]} (a) for $J_2/J_1=0.2$. 
As $J_2/J_1$ increases further, open surfaces emerge as shown in Figs.~\ref{fig:J1_J2_LT_wavevector[1]} (b), (c), and (d). 
Notice that the open surface at $J_2/J_1=0.5$ has a touching point at $\Gamma$, which means that 
collinear ordered phase with ${\bf q}=0$ is a part of the solutions. 
It is also interesting that open surfaces have distinct topologies for $0.2 \lesssim J_2/J_1 < 0.5$ 
and $ J_2/J_1 >0.5$. These behaviors stem from a special property of the hyperhoneycomb lattice structure. 
It will be shown later that this special property is related to the emergence of collinear stripy order 
near $J_2/J_1=0.5$ upon including quantum fluctuations. This will be discussed in details in 
Sec.~\ref{subsec:single-q} and Sec.~\ref{sec:semi-classical}. 

Now we determine the true classical ground-state manifold by examining which solutions on
the degenerate surfaces would satisfy the hard spin constraint. 
Careful investigations reveal that only special lines of the ordering wavevectors 
on the degenerate surfaces satisfy the hard spin constraint. 
Black lines in Figs.~\ref{fig:J1_J2_LT_wavevector[1]} (a)-(d) represent the ordering wavevectors that
meet the hard spin constraint for different values of $J_2/J_1$. 
In general, the degenerate lines lie on the two planes 
$\Gamma{\rm Z}{\rm T}{\rm Y}$ and $\Gamma {\rm X}{\rm A}{\rm Z}$ 
where the $\Gamma$, ${\rm Z}$, ${\rm T}$, ${\rm Y}$, ${\rm X}$, and ${\rm A}$ respectively indicate the 
high symmetry points in the Brillouin zone (See Appendix~\ref{app-hyperhoneycomb} for details).
When degenerate surfaces intersect with the Brillouin zone boundary, however, 
there are extra degenerate lines at the zone boundary perpendicular to the ${\hat {\bs q}}_z$ axis
as shown in Figs.\ref{fig:J1_J2_LT_wavevector[1]} (b), (c) and (d). 
These solutions with degenerate lines are shown to be consistent with the results of 
the single-${\bf Q}$ variational approach as discussed in the next section.

\subsection{Single-${\bf Q}$ variational approach}
\label{subsec:single-q}

As an alternative approach, we use a variational ansatz for the classical solutions that
satisfy the hard spin constraint explicitly. 
Treating the spins as classical three-component unit-vectors, 
we consider the following single-{\bf Q} variational ansatz.
\bea
{\bs S}_i = {\cal R}e [ {\bf d }    ~ e^{i ( {\bf Q} \cdot {\bf r}_i + \varphi_{s} ) }], \hspace{0.1in} 
{\bf d} =  \hat{{\bf e}}_1 + i {\hat{\bf e}}_2 ,
\label{eq:sQ-1}
\eea
with variational parameters ${\bf Q}$ and $\varphi_{s}$, where 
${\bf Q}$ is the ordering wavevector and $\varphi_s$ is a sublattice-$s$-dependent phase factor.
Here, $\hat{\bf e}_{1}$ and $\hat{\bf e}_2$ are orthogonal unit vectors and 
they can be freely chosen as the underlying Hamiltonian is SU(2) invariant. 
Using the variational ansatz in Eq.~\eqref{eq:sQ-1}, we find the parameter values of 
${\bf Q}$ and $\varphi_s$ that minimize the energy of 
the $J_1$-$J_2$ Heisenberg spin Hamiltonian ${\cal H}$ [Eq.~\eqref{eq:1}]. 
This variational ansatz always satisfies the hard spin constraint $| {\bf S}_i | =1$, 
but works only when the magnetic ordering is described by a single wavevector.
Black lines in Figs.~\ref{fig:J1_J2_LT_wavevector[1]} (a)-(d) show the wavevectors ${\bf Q}$,
which minimize ${\cal H}$ [Eq.~\eqref{eq:1}] for different values of $J_2/J_1$. 
We note that these single-${\bf Q}$ solutions are fully consistent with the ones of the Luttinger-Tisza analysis
supplemented with the hard spin constraint. 
Below we point out a number of important characteristics of these classical spin states that form
the line degeneracies.

(i) Magnetic frustration induced by the competition between $J_1$ and $J_2$ in the hyperhoneycomb
lattice leads to a manifold of {\it ``spiral line states"}, where degenerate ordering wavevectors form lines 
in the 3D Brillouin zone. It is interesting to compare this line degeneracy with
the classical ground states of the same Heisenberg model on the 3D diamond lattice,
where the ordering wavevectors form degenerate surfaces in the Brillouin zone or 
represent {\it ``spiral surface states"}.
\cite{bergman2007order,PhysRevB.78.144417}
The diamond lattice contains two sublattices in the unit cell and 
the nearest-neighbors (next-nearest-neighbors) connect different (same) sublattices.
When $J_1>0$ and $J_2=0$, the ground state is clearly the N\'eel state 
with the ordering wavevector ${\bf Q}=0$. 
In the case of $J_1=0$ and $J_2>0$, on the other hand, magnetic frustration is present and
degenerate classical spin states are characterized by ordering wavevectors aligned 
along any of three principal axes ${\bf Q} \sslash [100], [010], [001]$. Notice that all three
principal directions in three-dimensions are equally allowed in this degenerate manifold.
It is, therefore, natural to expect that the degenerate wavevectors of spiral states in 
the presence of both $J_1$ and $J_2$ would not exclusively occur on any particular plane 
or along a particular direction and rather form a degenerate surface.
In contrast, the next-nearest-neighbors on the hyperhoneycomb lattice are connected between not only the same sublattices but also different ones. When $J_2 >0$ and $J_1=0$, there
exists magnetic frustration as in the case of the diamond lattice. However,
the different connectivity for the next-nearest-neighbors leads to the degenerate wavevectors forming 
a circle on the $\Gamma X A_1 Y$ plane in the $J_2$-only model. 
In this case, the next-nearest-neighbor interactions already determine 
a special plane on which the degeneracy of the wavevectors resides.
Hence, one would expect that the degenerate wavevectors in the presence of
both $J_1$ and $J_2$ would not form a surface spanning all three directions in the
Brillouin zone and rather form degenerate lines/circles extended along two directions 
in a plane.

For the SU(2) invariant systems, the spiral plane on which the spins lie, can be freely chosen.
In crystals, however, this SU(2) symmetry can be easily broken by the crystal lattice potential 
combined with the spin-orbit coupling. The spin-orbit effect couples the spin and spatial rotations
allowed by the lattice symmetry. This may induce magnetic anisotropies and lock the spiral plane 
to be pointing along a special direction, depending on the ordering wavevectors.
Notably, iridium electrons have strong spin-orbit coupling 
and certain magnetic anisotropies are likely to be present. 
In Sec.~\ref{sec:discussion}, we discuss how such anisotropy effects select 
a particular spiral plane, depending on the ordering wavevectors.

(ii) At $J_2/J_1 = 0.5$, the line degeneracy of the classical ground state manifold 
contains not only spiral ordered phases, but also the collinear stripy phase with the ordering 
wavevector ${\bf Q}=0$, where the spins in sublattices \{0, 1\} and \{2, 3\} 
point in opposite directions to each other 
(See Fig.~\ref{fig:spin_config} (c) for a schematic picture of the stripy order).
When $J_2/J_1 =0.5$, the ratio of $J_2/J_1$ becomes exactly the same as the ratio of 
the number of nearest neighbor bonds and that of next-nearest neighbor 
bonds, which are three and six respectively.
The stripy order allows only 2/3 of the nearest-neighbor bonds to gain the
antiferromagnetic spin exchange energy. However, it also gains the antiferromagnetic spin exchange energy for 
2/3 of the next-nearest-neighbor bonds. As a result, the bond energies associated with $J_1$ and $J_2$ become equal at $J_2/J_1=0.5$,
making the energy of the collinear stripy phase degenerate with those of competing spiral ordered phases.
%
In the following section (Sec.~\ref{sec:semi-classical}), we explore the effects of quantum fluctuations 
using two different approaches: the $1/S$ expansion in the linear spin-wave theory and Schwinger boson
mean-field method. In particular, we show that the stripy order wins over the competing spiral ordered phases 
in a finite range of parameters near $J_2/J_1 = 0.5$ upon including quantum fluctuations, 
despite the fact that the stripy phase is a part of the classical ground state manifold only at $J_2/J_1 = 0.5$.

\section{Quantum order-by-disorder, emergence of stripy phase, and quantum spin liquids}
\label{sec:semi-classical}

Quantum fluctuation may lift the line degeneracy in the classical-ground-state 
manifold, identified in the previous section. This {\sl quantum order-by-disorder} effect
may select certain magnetically ordered phase among the degenerate ground states.
When quantum fluctuations become extremely strong, however, magnetic ordering 
would be completely suppressed and various quantum spin liquid phases may
become emergent ground states. Here we explore both possibilities using two
different approaches.

In Sec.~\ref{subsec:HP-bosons}, we first use the large-$S$ anaylsis of the linear spin wave theory 
to investigate quantum order-by-disorder effects by computing zero-point quantum-fluctuation 
energy of degenerate classical ground states. 
We show that, in general, quantum fluctuations select certain
magnetically ordered phases with ordering wavevectors lying along the high 
symmetry directions in the Brillouin zone. It is also found that quantum fluctuations favor
collinear ordered states such as the N\'eel and stripy phases in a much wider
region of the parameter space, compared to the classical limit. 
In Sec.~\ref{subsec:schwinger-bosons}, it is shown that the Schwinger boson approach 
results in similar quantum order-by-disorder effects. We also explore emergent quantum
spin liquid phases in the Schwinger boson mean-field theory when quantum fluctuations 
are very strong. 

\subsection{Large-$S$ analysis}
\label{subsec:HP-bosons}

We now consider the linear spin-wave theory via the Holstein-Primakoff boson representation.
In order to include quantum fluctuations, we adopt the following spin-coordinate frame at each site $i$.
\bea
\hat{\bs{z}}_i &=& \hat{S}_i^{\rm{cl} } = {\cal R}e [ \bs{d} ~ e^{i (\bs{Q} \cdot {r_i} + \varphi_s)} ], \nn \\
\hat{\bs{x}}_i &=& - {\cal I}m [  \bs{d} ~e^{i (\bs{Q} \cdot {r_i} +\varphi_s) } ], \nn \\
\hat{\bs{y}} &=&  \frac{i}{2} \bs{d} \times \bs{d}^* = \hat{\bs{e}}_3,
\label{eq:subsec-hp-1}
\eea
where the local $\hat{\bs{z}}_i$ axis is defined to be parallel to the direction of the classical spin order 
[See Eq.\eqref{eq:sQ-1}].
For coplanar spiral states, one of the coordinate axis, $\hat{\bs{y}}$, can be taken as the
normal vector of the spiral plane and it would be site-independent. 
In the large-$S$ limit, the linearized Holstein-Primakoff transformation can be written as
\be
\bs{S}_i = (S-n_i) \hat{\bs{z}}_i + \sqrt{2S} \Big( a^\dagger_i \frac{ \hat{\bs{x}}_i  + i \hat{\bs{y}}}{2} 
+ a_i  \frac{\hat{\bs{x}}_i  - i \hat{\bs{y}}}{2} \Big),
\label{eq:subsec-hp-2}
\ee 
where $S$ is the spin magnitude, 
$a_i^\dagger (a_i)$ are boson creation (annihilation) operators, and $n_i = a_i^\dagger a_i$ is the boson density operator. 
Using Eq.\eqref{eq:subsec-hp-2}, we expand $\mathcal{H}$ in Eq.\eqref{eq:1} 
up to the quadratic order of boson operators, which results in the leading order Hamiltonian
in the $1/S$ expansion. 
We can now easily evaluate the zero-point quantum fluctuation energy of the Holstein-Primakoff bosons
(For more details, see Appendix~\ref{app:HP-bosons}). 
One finds that quantum fluctuations select particular ordering wave vectors 
along the high symmetry lines of the Brillouin zone and 
such ordering wave vectors are depicted as green spheres in Figs.~\ref{fig:J1_J2_LT_wavevector[1]} (a)-(d) for 
different values of $J_2/J_1=0.2,0.3, 0.5$ and $0.7$. 
Depending on parameter regions, different kinds of magnetically ordered phases are selected as follows.
\begin{itemize}
\item{{\it $\Gamma$Z spiral} : For $ 0.17 \lesssim J_2/J_1 < 0.5$, quantum fluctuations select 
the ordering wavevectors {${\bf Q}=\pm q(1,1,0)$ along the $\Gamma$-Z line}. 
Some examples are shown in Figs.~\ref{fig:J1_J2_LT_wavevector[1]} (a) and (b) for
$J_2/J_1 =0.2$ and $J_2/J_1=0.3$, respectively.}
\item{{\it $\Gamma$XY spiral} : For $ J_2/J_1 >0.5 $, quantum fluctuations favor
the ordering wavevectors {${\bf Q}=\pm q(1,-1,0)$ along the $\Gamma$-X line} 
and {${\bf Q}=\pm q(0,0,1)$ along the $\Gamma$-Y line}.
Figure~\ref{fig:J1_J2_LT_wavevector[1]} (d) illustrates such selection of the ordering wavevectors 
for $J_2/J_1 = 0.7$.}
\end{itemize}

%
\begin{figure}[t]
\scalebox{0.65}{\includegraphics{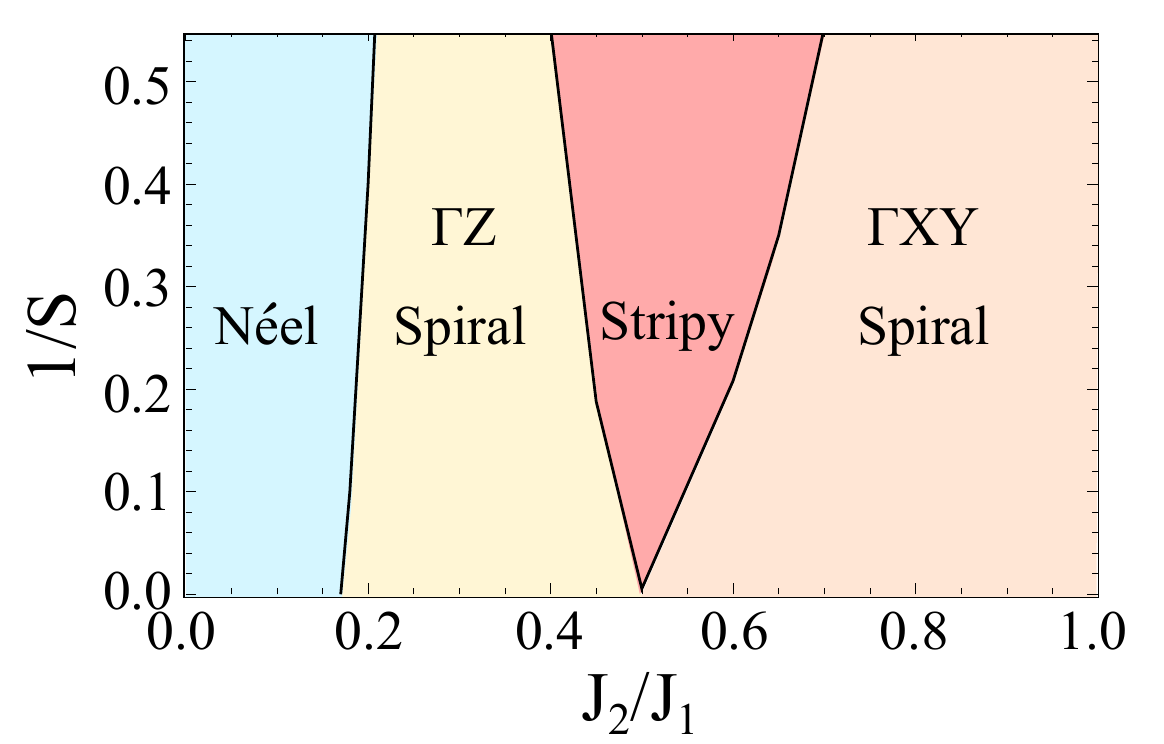}}
\caption{(Color online) Magnetic phase diagram of the $J_1$-$J_2$ Heisenberg model on the hyperhoneycomb lattice 
as a function of inverse spin magnitude $1/S$ and the ratio $J_2/J_1$, where zero-point quantum fluctuations
are included in the linear spin-wave theory. 
Notice that the collinear orders such as the stripy and N\'eel phases become the ground states for 
a wider region of parameter space as quantum fluctuations become stronger or $1/S$ becomes bigger.
}
\label{fig:subsec-hp-1}
\end{figure}

Figure~\ref{fig:subsec-hp-1} shows the phase diagram as a function of inverse spin-magnitude $1/S$ 
and the ratio $J_2/J_1$ upon including zero-point quantum-fluctuation energy corrections. 
Notice that the collinear magnetic orders such as N\'eel and stripy phases,
win over spiral orders in a wider range of parameter space as quantum fluctuations become stronger
or $1/S$ becomes bigger.
This is reminiscent of the same trend found in previous studies 
on various systems.\cite{SovPhys,PhysRevLett.62.2056,PhysRevLett.102.137201}
For example, in the classical limit $S \rightarrow \infty$, the stripy phase is a part of the classical-ground-state 
manifold only at a single point $J_2/J_1 =0.5$ (See Sec.~\ref{subsec:single-q} for details).
However, the collinear stripy phase becomes the ground state in a wider
region of parameter space near $J_2/J_1 \sim 0.5$ when $S$ becomes smaller.
The same trend exists for the N\'eel phase, leading to a wider region of N\'eel order 
near $J_2/J_1 \sim 0.17$. 
Of course, the computations of zero-point quantum fluctuations in the linear
spin-wave theory are valid only for large spin magnitude $S$ or small $1/S$. 
In order to access the strong quantum fluctuation regime, we now turn to 
the Schwinger boson analysis that can be used to study both semi-classical
and strongly quantum regimes on an equal footing.

\begin{figure}
 \centering
 \includegraphics[width=0.4\linewidth,angle=270]{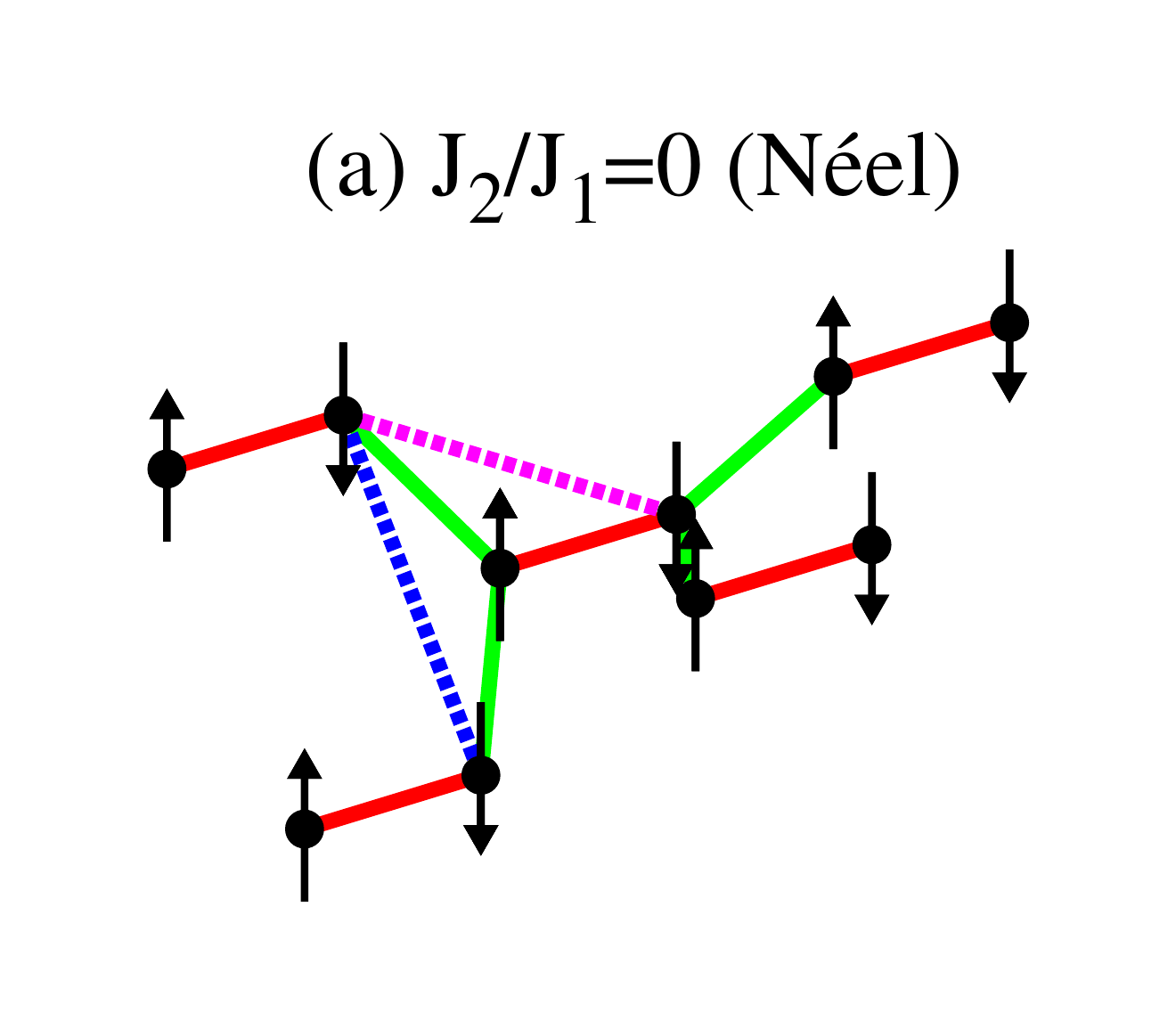}
 \includegraphics[width=0.4\linewidth,angle=270]{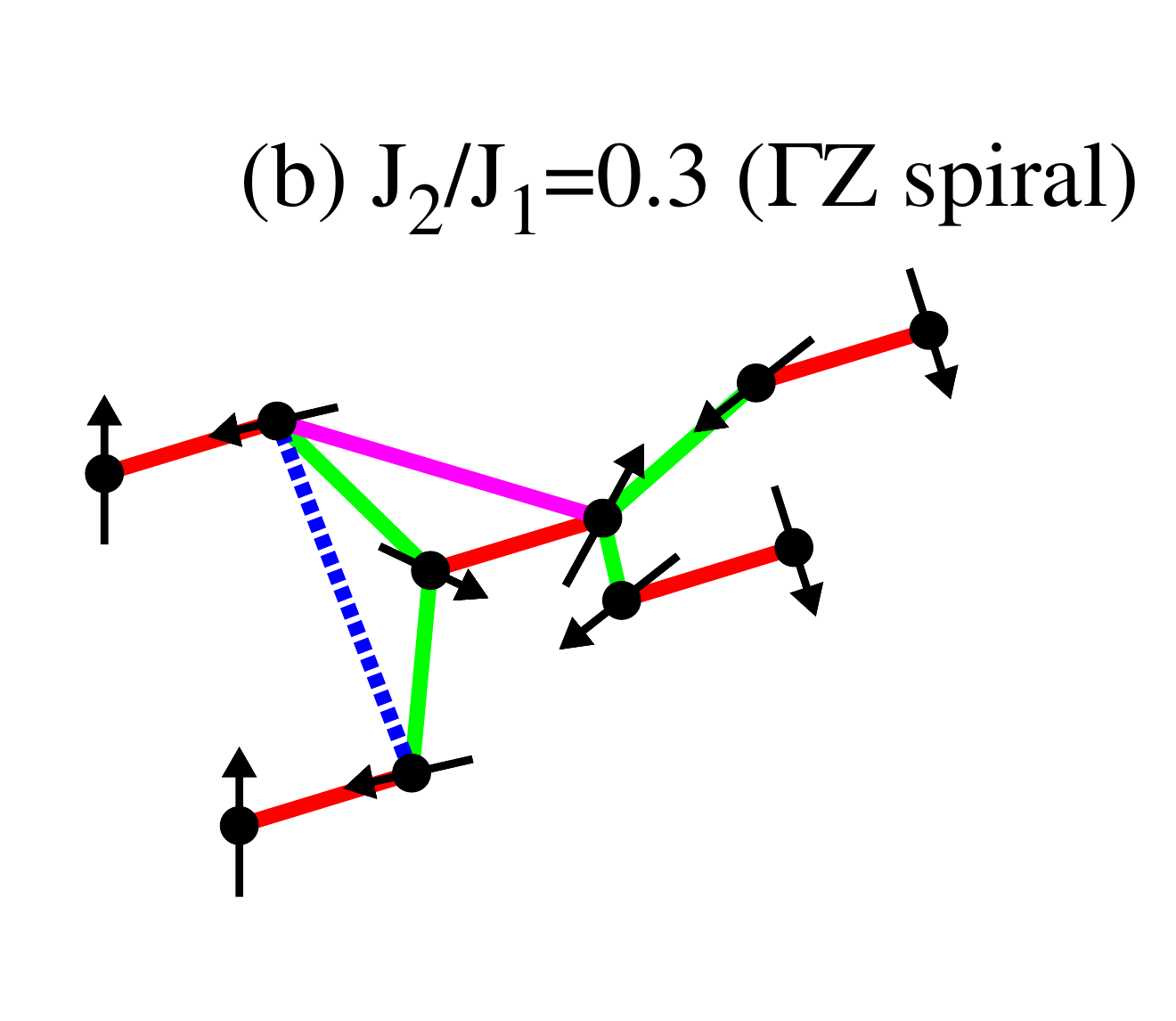}
 \includegraphics[width=0.4\linewidth,angle=270]{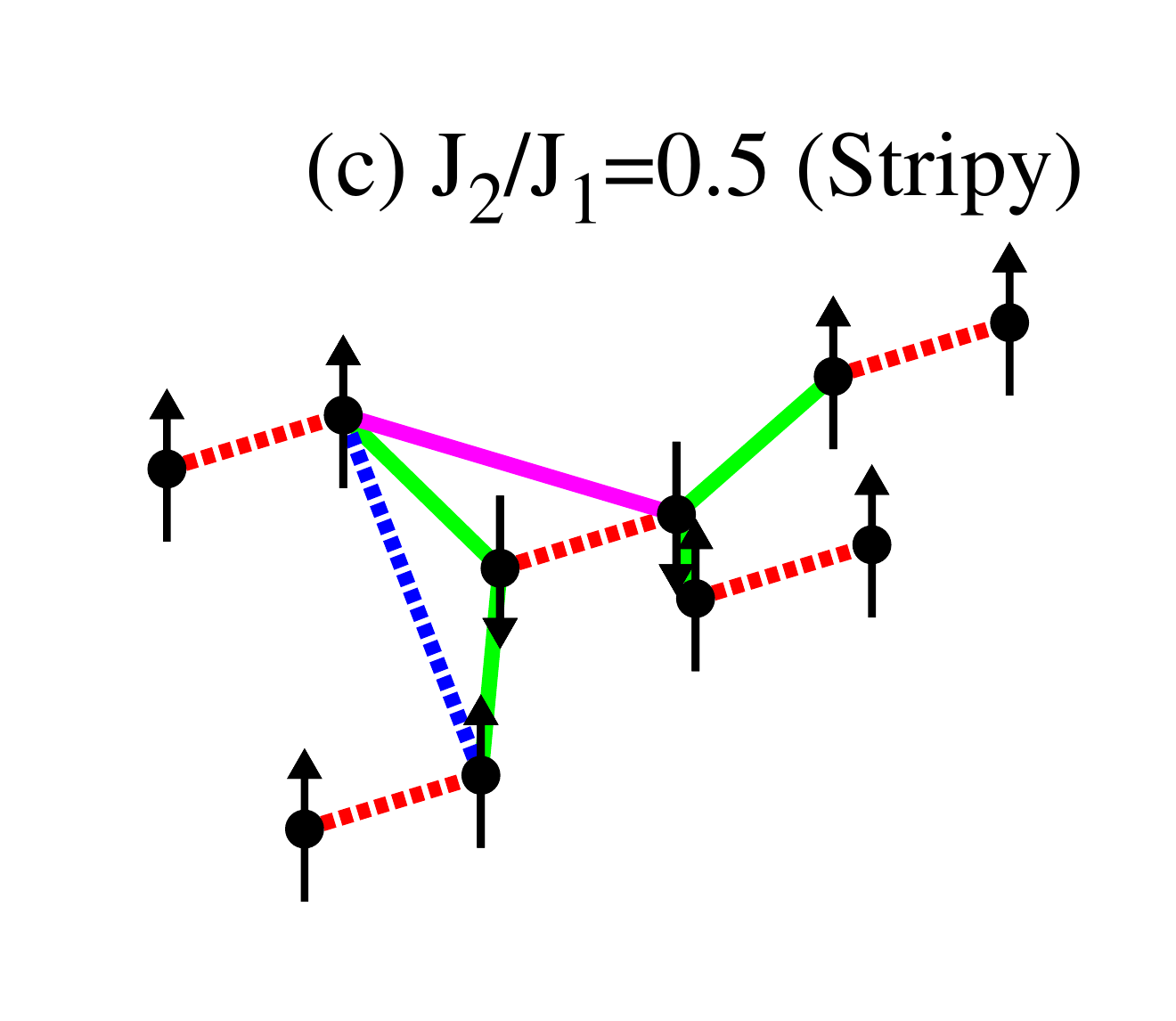}
 \includegraphics[width=0.4\linewidth,angle=270]{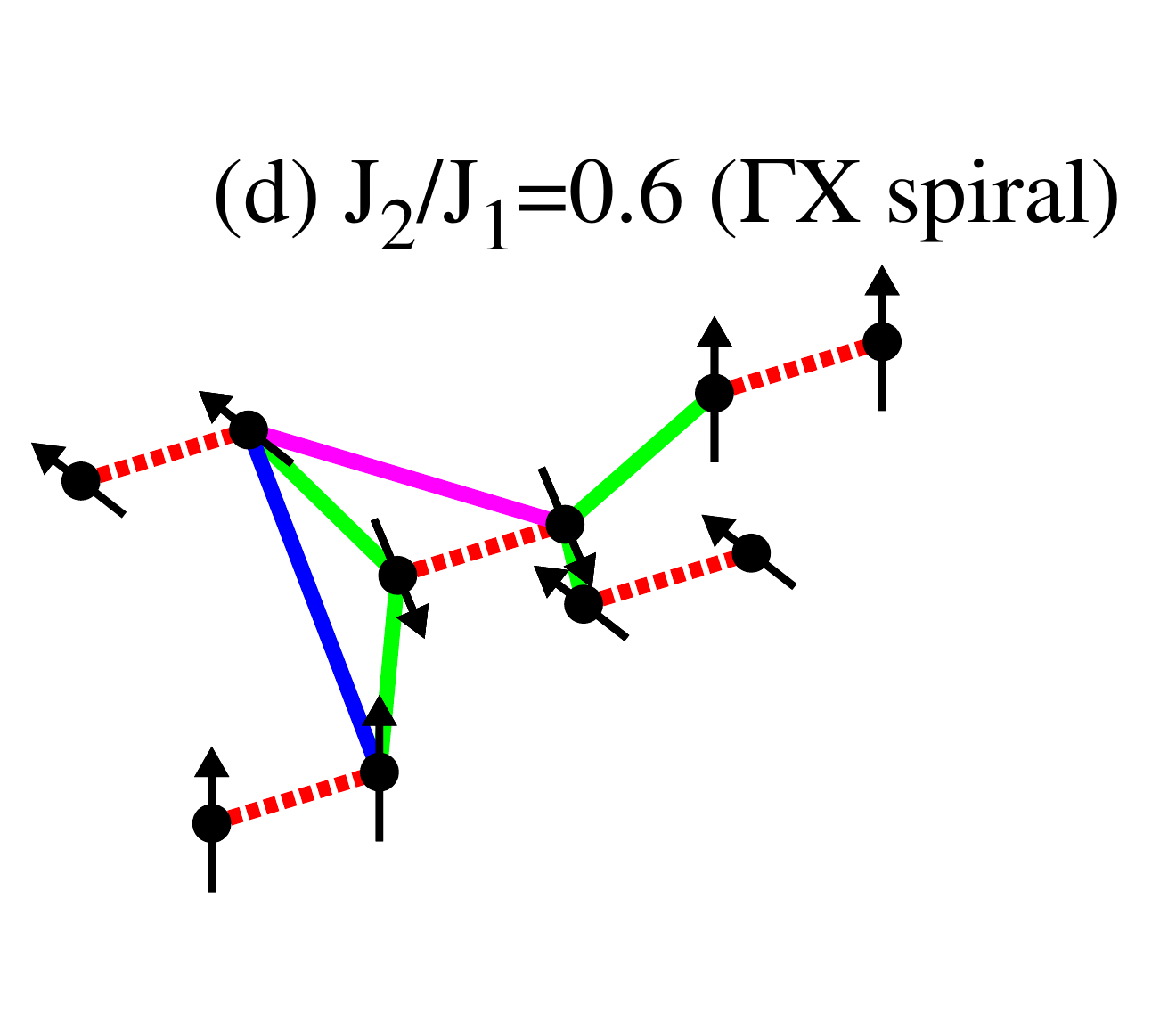}
 \caption{(Color online) Classical spin configurations and Schwinger boson mean-field parameters. Spin moments are represented by arrows. Mean-field parameters $\eta_{n}~(n=1,2,3,4)$ in the Schwinger boson theory are denoted by red, green, blue, and magenta lines, respectively. Solid lines corresponds to antiferromagnetic correlations ($\eta_n > 0$) between two sites and dashed lines represent ferromagnetic arrangements ($\eta_n = 0$). For clarity, we show $\eta_{3,4}$ only on two $J_2$-bonds in each figure with the rest of them being omitted.
 \label{fig:spin_config}}
\end{figure}

\subsection{Schwinger boson approach}
\label{subsec:schwinger-bosons}

\begin{figure}[t]
 \centering
 \includegraphics[width=1\linewidth]{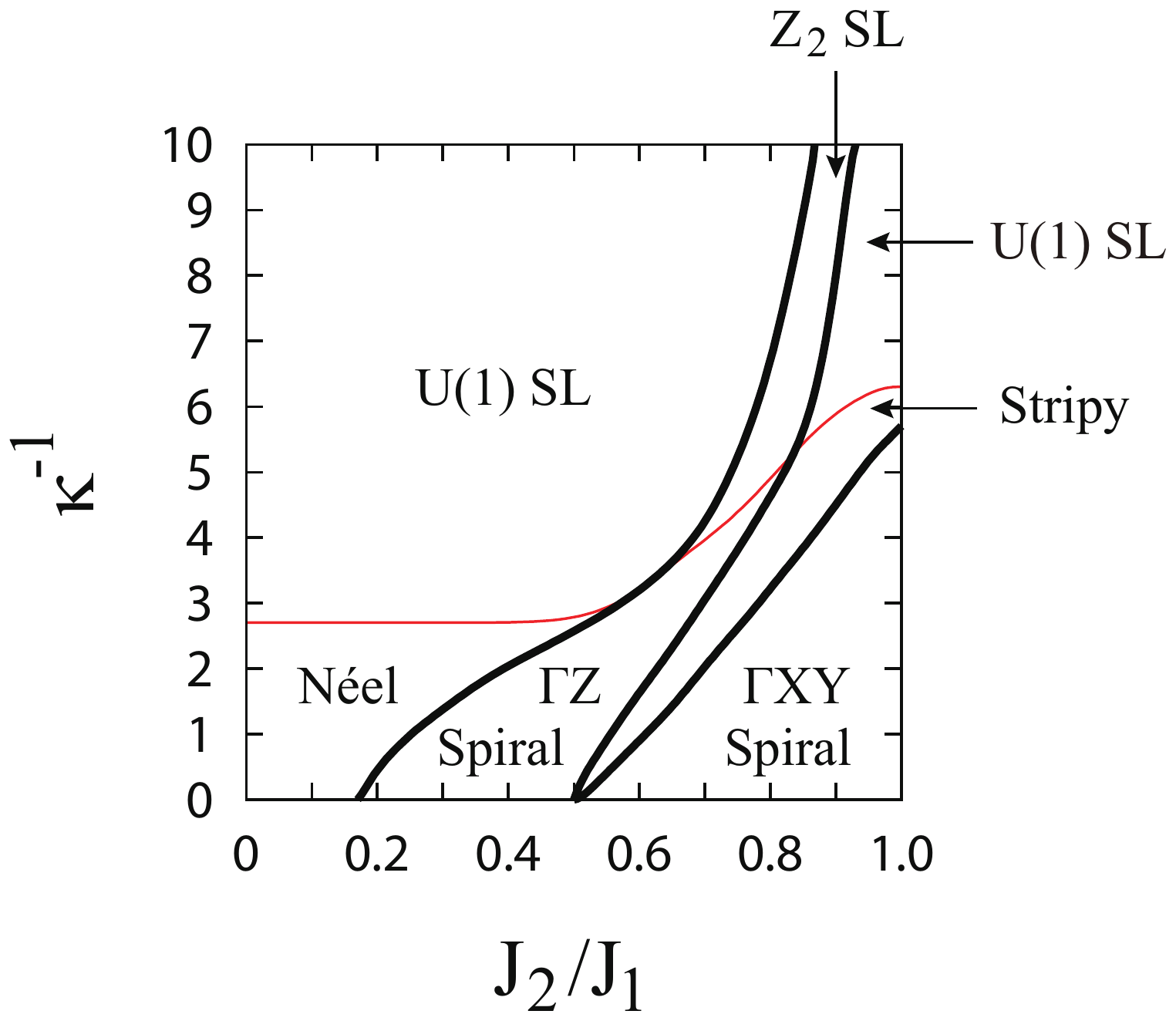}
  \caption{(Color online) Mean-field phase diagram in the Schwinger boson theory. 
At small $1/\kappa$, quantum order-by-disorder effects select the Neel, $\Gamma$Z spiral, stripy, and $\Gamma$XY spiral phases.
When quantum fluctuations become stronger (or $1/\kappa$ becomes bigger), U(1) and Z$_2$ quantum spin liquid phases arise.
Thick solid (red thin) lines represent first (second) order phase transitions.
 \label{fig:SBMFT_PD}}
\end{figure}

In the Schwinger boson theory\cite{1988_arovas,PhysRevLett.66.1773,PhysRevB.45.12377}, 
the spin operator is represented in terms of spin-carrying bosons, $b_{i\alpha}$:
\begin{equation}
 {\bf S}_i = \frac{1}{2} b_{i\alpha}^{\dagger} \boldsymbol{\sigma}_{\alpha\beta} b_{i\beta},
 \label{eq:sb_rep}
\end{equation}
where $\boldsymbol{\sigma}_{\alpha\beta}$ are the Pauli matrices ($\alpha,\beta\in\{\uparrow,\downarrow\}$), and summations over repeated Greek indices are assumed. 
Here the boson density at each site is related to the magnitude of the spin $S$ via
\begin{equation}
 n_{b} = b_{i\alpha}^{\dagger} b_{i\alpha} = \kappa,
 \label{eq:constraint}
\end{equation}
where $\kappa=2S$.
Using the Schwinger boson representation, we consider the following mean-field Hamiltonian for the $J_1$-$J_2$ Heisenberg model in Eq. (\ref{eq:1}).
\begin{eqnarray}
 \mathcal{H}_{MF} 
 &=&
 \sum_{i>j} \frac{J_{ij}}{2} \left( \left| \eta_{ij} \right|^2 - \eta_{ij} b_{i\alpha}^{\dagger} \epsilon_{\alpha\beta} b_{j\beta}^{\dagger} + \textup{H.c.} \right)
 \nonumber\\
 &+&
 \sum_{i} \lambda_i \left( b_{i\alpha}^{\dagger} b_{i\alpha} - \kappa \right),
 \label{eq:H_MF}
\end{eqnarray}
where ${\eta}_{ij} = \langle b_{i\alpha} \epsilon_{\alpha\beta} b_{j\beta} \rangle$ is the mean-field parameter ($\epsilon_{\alpha\beta}$ is the antisymmetric tensor) and $\lambda_i$ is the Lagrange multiplier to implement the constraint on spin magnitude in Eq. (\ref{eq:constraint}). 
It has been known that such mean-field solutions become exact in the large-$N$ limit of the Sp(N) generalized model, where $N$ flavors of 
bosons, $b_{i \alpha m} (m = 1, 2, ..., N)$, are introduced and the constraint is generalized to 
$n_b = b^{\dagger}_{i \alpha m} b_{i \alpha m} = \kappa N$. The large-$N$ limit is then taken by fixing $n_b/N = \kappa$.
Notice that $\kappa$ in this limit plays the role of $2S$ in the SU(2) case, namely the large(small)-$\kappa$ limit corresponds
to semi-classical (quantum) regime. Thus the mean-field solution is {\sl non-perturbative} in $\kappa$ or $2S$
in contrast to the spin-wave theory.
Here we directly work with the mean-field solutions in the SU(2) limit.
In this formulation, the bose condensation at large-$\kappa$ leads to magnetically ordered phases while 
quantum spin liquid phases with gapped spin-carrying bosons, dubbed spinons, appear at small-$\kappa$.
We explicitly include the condensate $x_{i\alpha}=\langle b_{i\alpha} \rangle$ degrees of freedom in Eq.\eqref{eq:H_MF} and 
minimize the energy $\langle \mathcal{H}_{MF} \rangle$ with respect to $\eta_{ij}$, $\lambda_i$, and $x_{i\alpha}$.

\subsubsection{Classical limit and mean-field ansatz}

The classical limit can be obtained by taking $\kappa \rightarrow \infty$ in the Schwinger boson mean-field theory,
where the classical spins ${\bf S}^c_i$ and mean-field link variables $\eta_{ij}^c$ are written
in terms of condensate amplitudes of bosons:
\begin{eqnarray}
{\bf S}_i^c&=&\frac{1}{\kappa}x_{i\alpha}^*\boldsymbol{\sigma}_{\alpha\beta}x_{i\beta},
\label{eq:sb-9}
\\
\eta_{ij}^c&=&\frac{1}{\kappa}x_{i\alpha} \epsilon_{\alpha\beta} x_{j\beta},
\label{eq:sb-10}
\end{eqnarray}
where the scaled ${\bf S}_i$ and $\eta_{ij}$, normalized by the boson density $\kappa$, satisfy 
$|{\bf S}_i^c|=1$ and ${\bf S}_i^c \cdot {\bf S}_j^c = - 2 |\eta_{ij}^c|^2+1$. 
Comparing Eqs.~\eqref{eq:sb-9} and \eqref{eq:sb-10} with the single-{\bf Q} variational ansatz for the
spiral spin states in Eq.~\eqref{eq:sQ-1}, one can determine the corresponding expressions of $x_{i\alpha}$ and 
$\eta_{ij}$ for the degenerate classical ground states investigated earlier. 
Upon including quantum fluctuations when $\kappa $ is large and finite, we find that quantum order-by-disorder chooses the
same set of magnetically ordered phases, namely N\'eel, Stripy, $\Gamma Z$, $\Gamma XY$ spiral orders.
It can be shown that the mean-field ansatz for these selected states requires four independent parameters 
$\eta_n$ ($n=1,\cdots,4$). 
Figures~\ref{fig:spin_config} (a)-(d) illustrate four different classical spin states: N\'eel, $\Gamma Z$ spiral, 
stripy and $\Gamma X$ spiral states. The corresponding mean-field parameters $\eta_n$ ($n=1,2,3,4$) are
also shown in the same figures.
Colored solid lines indicate four different mean-field parameters $\eta_n$ ($n=1,2,3,4$) for antiferromagnetic correlations 
and colored dashed lines represent ferromagnetic arrangement between two sites. The parameters $\eta_{1,2}$ are defined for the nearest-neighbor $J_1$ bonds and $\eta_{3,4}$ are for 
the next-nearest-neighbor $J_2$ bonds (see Appendix \ref{app:sbmft} for more details). 
As quantum fluctuations increase as $\kappa$ becomes smaller, magnetically ordered phases are 
suppressed and the corresponding condensate densities of bosons vanish. This marks the 
transition to quantum spin liquid phases. In the next section, we explore the resulting phase diagram for
various values of $\kappa$ and $J_2/J_1$.

\subsubsection{Mean-field phase diagram}

\begin{figure}
 \centering
 \includegraphics[width=0.8\linewidth,angle=270]{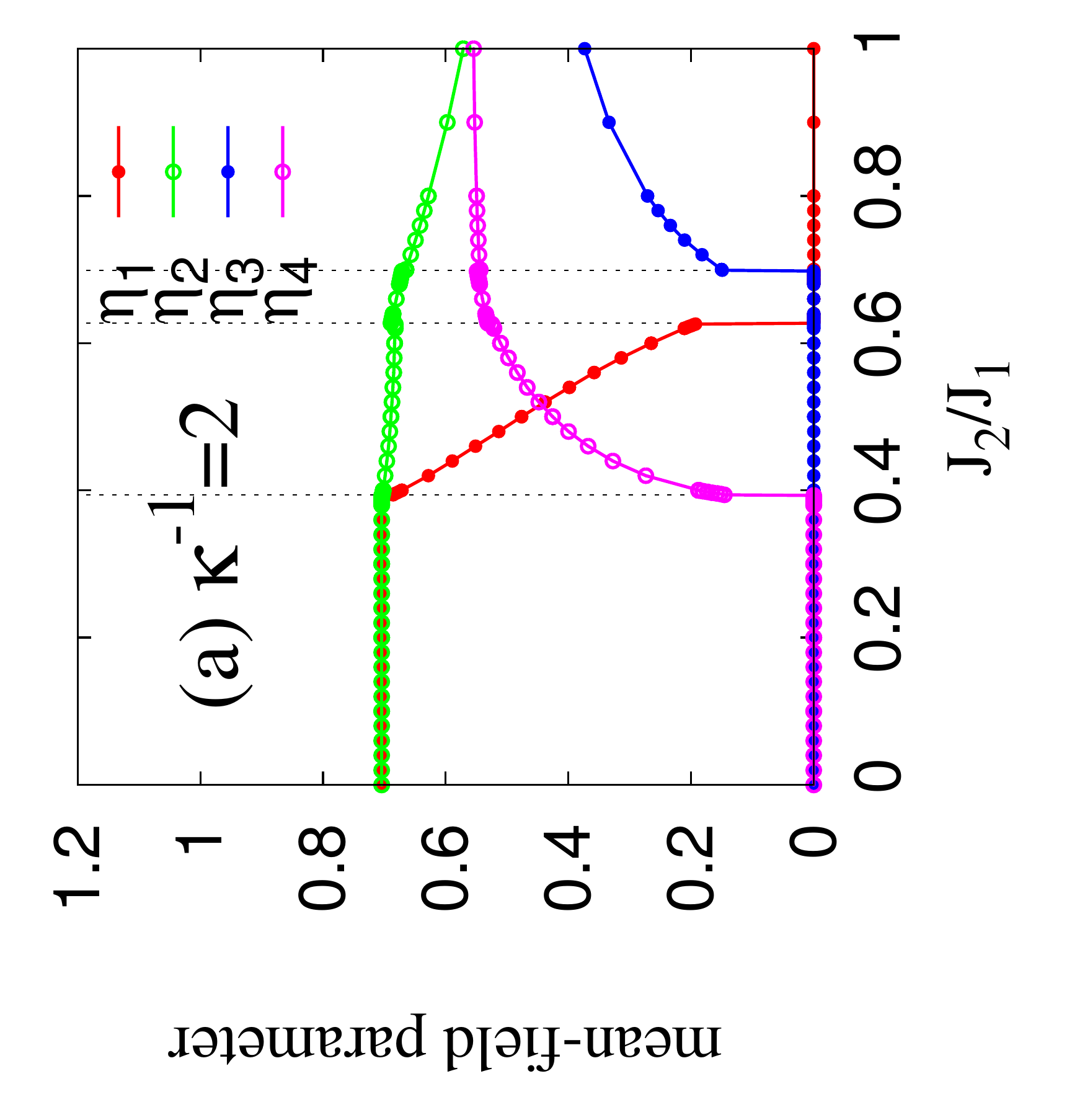}
 \includegraphics[width=0.8\linewidth,angle=270]{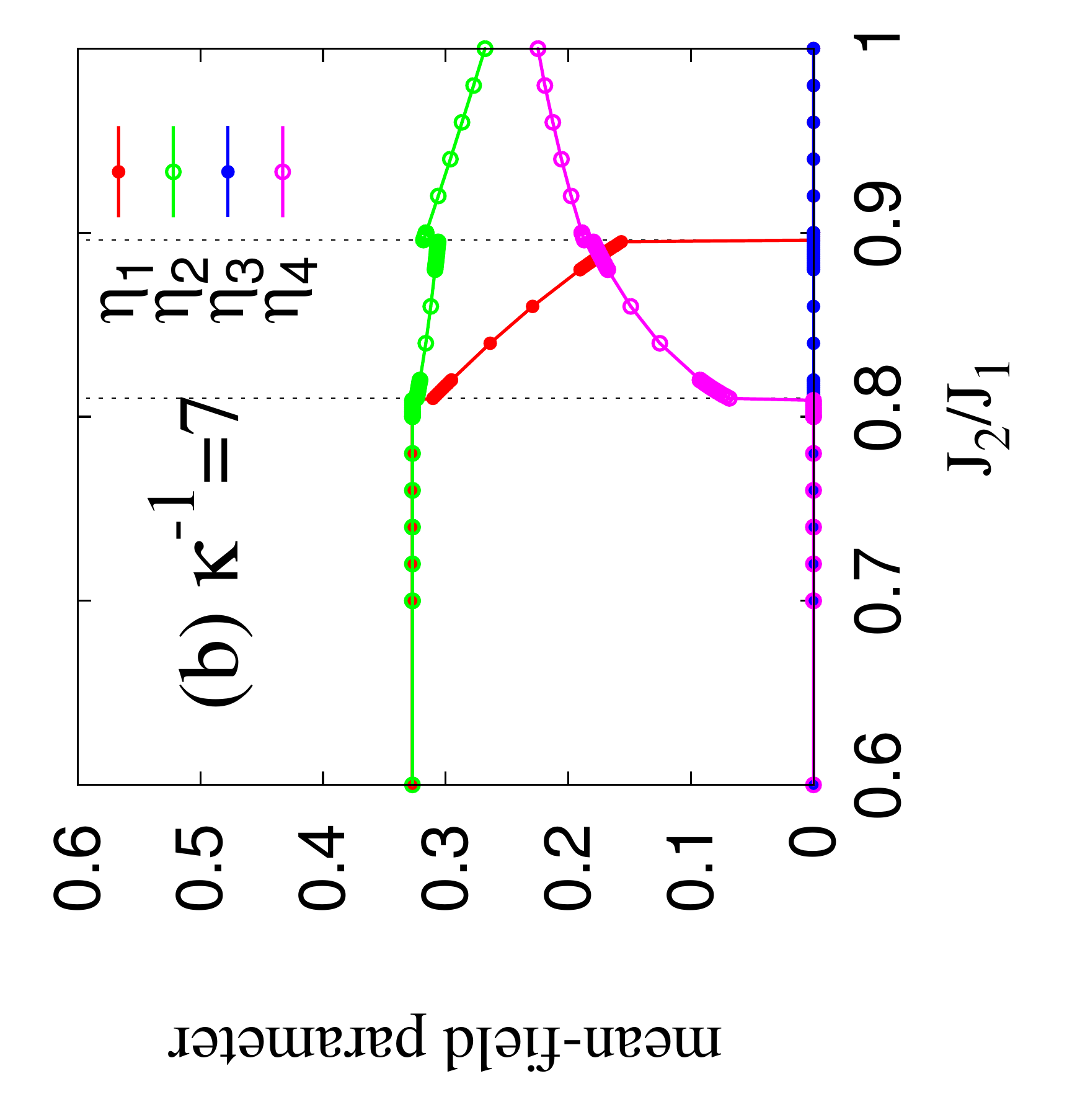}
 \caption{(Color online) Mean-field parameters, $\{ \eta_1, \eta_2, \eta_3, \eta_4 \}$, as functions of $J_2/J_1$ at $1/\kappa=2$ (magnetic order regime) and $1/\kappa=7$ (spin liquid regime). }
 \label{fig:MF}
 \end{figure}

Figure~\ref{fig:SBMFT_PD} shows the mean-field phase diagram as a function of $1/\kappa$ and 
the ratio of $J_2/J_1$. 
In the limit $\kappa \rightarrow \infty$, it successfully recovers all the classical magnetic phases 
with the phase boundaries consistent with the classical phase diagram. 
As an example, the behaviors of mean-field parameters as a function of $J_2/J_1$ for $1/\kappa = 2$ 
are shown in Fig.~\ref{fig:MF} (a).
As shown in Fig.~\ref{fig:spin_config}, the classical magnetic orders are described by 
different mean-field structures for various ranges of $J_2/J_1$: 
(i) $\eta_{1,2}>0$ for the N\'eel, (ii) $\eta_{1,2,4}>0$ for the $\Gamma Z$ spiral, 
(iii) $\eta_{2,4}>0$ for the stripy, and (iv) $\eta_{2,3,4}>0$ for the $\Gamma XY$ spiral phases. 

Just like the results of the linear spin-wave theory, the collinear orders such as the N\'eel and stripy phases exist
in a wider region of the parameter space as quantum fluctuations become stronger or $1/\kappa$ becomes bigger.
On the other hand, the phase boundaries between different phases look somewhat different, 
and this happens
because quantum fluctuations enter differently in the linear spin-wave theory and Schwinger boson approach.
At the mean-field level, the phase transitions between magnetically ordered phases at finite $1/\kappa$ 
are first order (black thick solid lines) as is evident from the behaviors of the mean-field parameters shown in Fig.~\ref{fig:MF}.

When quantum fluctuations are further increased, second order phase transitions from the magnetically ordered phases
to quantum spin liquid states occur at some critical $(1/\kappa)_c$. Quantum spin liquid phases for $1/\kappa > (1/\kappa)_c$ possess
gapped bosonic spinon excitations. In the range of $0 \leq J_2/J_1 \leq 1$, we find $(1/\kappa)_c \simeq 3-6$.
U(1) and Z$_2$ spin liquid phases arise for $1/\kappa > (1/\kappa)_c$ upon quantum disordering the N\'eel/stripy and spiral orders.
As is well known, U(1) and Z$_2$ spin liquid phases typically occur via second order phase transition from collinear
and spiral ordered phases, respectively.
To be more specific, we show the behaviors of mean-field parameters for $1/\kappa = 7$ in Fig.~\ref{fig:MF} (b).
Here U(1) and Z$_2$ refer to the gauge structure of quantum spin liquid phases, which can be characterized
by physical quantities invariant under the corresponding gauge transformations. For example, the U(1) spin liquid arising from
the N\'eel state is distinct from the U(1) spin liquid associated with the stripy order as gauge invariant quantities, such as
the gauge-invariant products of link variables around closed loops,\cite{tchernyshyov2006flux} are different in two phases.  

\begin{figure}
 \centering
 \includegraphics[bb=50 50 560 240,clip,width=2.\linewidth,angle=270]{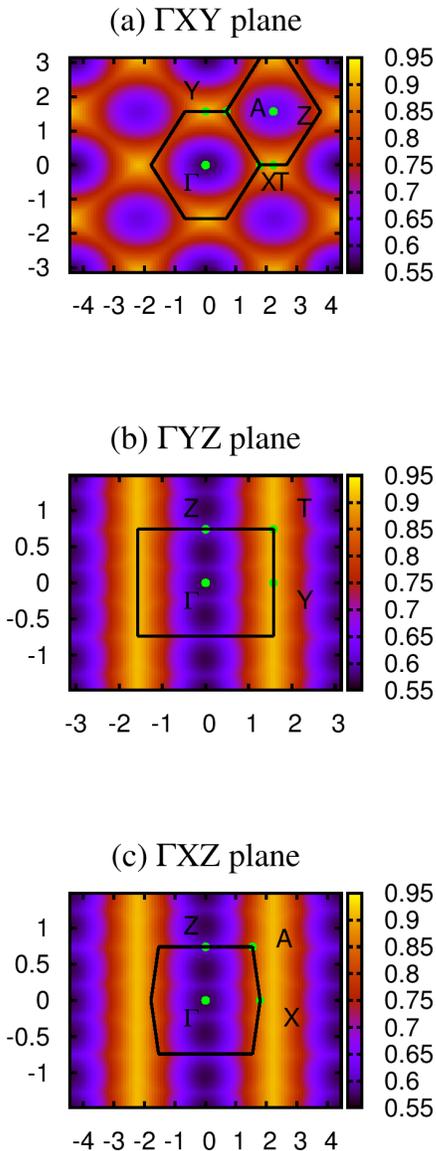}
 \caption{(Color online) Lower edge of the two-spinon spectrum in the Z$_2$ spin liquid. The spectrum is calculated at $J_2/J_1=0.84$ and $1/\kappa=7$ in the phase diagram of Fig.~\ref{fig:SBMFT_PD}. Three plots show the spectra in three different planes in the first Brillouin zone: the $\Gamma$XY, $\Gamma$YZ, and $\Gamma$XZ planes.
 \label{fig:two_spinon}}
\end{figure}

Notice that the Z$_2$ spin liquid that emerges from the $\Gamma Z$ spiral order (or the Z$_2$ spin liquid arising from 
the $\Gamma XY$ spiral, not shown in Fig.~\ref{fig:SBMFT_PD}) is different from the Z$_2$ spin liquid discovered in the Kitaev-Heisenberg model on the same lattice.
The former has gapped bosonic spinon excitations and the latter supports gapless Majorana fermion
excitations \cite{PhysRevB.89.045117,2013arXiv1309.1171K}.
In order to further characterize this Z$_2$ spin liquid, we show the dispersion of the lower edge of 
the (gapped) spinon-antispinon continuum in different planes of the Brillouin zone in Fig.~\ref{fig:two_spinon}, 
which can be obtained from $E_2({\bf Q})=\textup{min}_{\bf p} [\epsilon({\bf Q}-{\bf p})+\epsilon({\bf p})]$, where $\epsilon({\bf p})$ is the dispersion
of the gapped bosonic excitations. 
Figure~\ref{fig:two_spinon} corresponds to the case of $J_2/J_1=0.84$ and $1/\kappa=7$, and the dispersions are shown for three different planes 
in the first Brillouin zone: $\Gamma$XY, $\Gamma$YZ, and $\Gamma$XZ planes. 
The minimum energy of the two-spinon continuum occurs at ${\bf Q}=\pm q(1,1,0)$ along the $\Gamma$Z line, which is consistent with the fact 
that this Z$_2$ spin liquid emerges from the $\Gamma$Z spiral order.
These dispersions of the lower edge of the spinon-antispinon continuum can, in principle, be measured in
the spin structure factor via neutron scattering experiments.

\section{Discussion and Outlook}
\label{sec:discussion}

In this work, we investigated the nature of magnetic frustration and emergent quantum phases in the SU(2) symmetric $J_1$-$J_2$ Heisenberg model
on the hyperhoneycomb lattice.
We identified degenerate classical-ground-state manifold and studied the effects of quantum fluctuations. 
For $J_1 > 0$ and $J_2 > 0$, it was found that quantum order-by-disorder effects select spiral ordered phases 
as well as collinear stripy and N\'eel phases. 
Such collinear phases can be shown to be equivalent to zigzag and ferromagnetic phases when $J_1 < 0$ and $J_2 > 0$ via the transformation Eq.\eqref{eq:model-2}.
Possible quantum spin liquid phases in the strong-quantum-fluctuation regime were also identified.

In real materials such as $\beta$-Li$_2$IrO$_3$, magnetic anisotropies are likely to exist due to spin-orbit coupling.
Hence, it is useful to investigate what the effects of such anisotropies are on magnetically ordered phases 
studied in this paper. While the full study of anisotropic spin interactions is beyond the scope of this paper, 
here we simply focus on coplanar spiral ordered phases and study how magnetic anisotropies would
affect such phases. 
In coplanar spiral phases, the plane where the spins are lying would rotate with a pitch
consistent with the ordering wavevector. Such a spiral plane is not fixed in SU(2)-invariant systems and can 
be chosen freely. In the presence of magnetic anisotropies, however, the spiral plane may be locked to 
particular directions. 

We now consider the spiral phases discussed in previous sections: $\Gamma X$, $\Gamma Y$,
and $\Gamma Z$ spirals with ordering wavevectors lying along the corresponding high symmetry directions. 
In order to understand how the spiral plane may be constrained, we consider the Landau free energy as 
a function of ${\bf d}$ which is defined in Eq.~\eqref{eq:sQ-1}. 
When the system develops a spiral state with a specific ordering wavevector {\bf Q},  
the full lattice symmetries of the hyperhoneycomb lattice are broken and only a subset of lattice
symmetries remains. We use this subset of symmetries and construct the symmetry-invariant free
energy (Detailed analysis is shown in Appendix~\ref{app-symmetry}). For the ordering wavevectors lying along $\Gamma$-Z, 
$ \Gamma$-X and $\Gamma$-Y, 
we denote the corresponding free energies as $f_{\Gamma {\rm Z}}$, $f_{\Gamma {\rm X}}$ 
and $f_{\Gamma {\rm Y}}$, respectively. 
The free energies can be easily expressed in terms of a unit-vector $\hat{\bf e}_3$ 
normal to the spiral plane, {\it i.e.} $\hat{\bf e}_3 =( { e}_3^x , ~{e}_3^y ,~ {e}_3^z )  = \hat{\bf e}_1  \times \hat{ \bf e}_2 $:
\bea
f_{\Gamma {\rm Z}} &=& f_{\Gamma {\rm Y}} = 
c_1~ \Big(({e}_3^{x})^2  + ({e}_3^y)^2 \Big) + c_2 ~( { e}_3^x ~{e}_3^y ) \\
f_{\Gamma {\rm X} } &=& c_1 ~\Big(({e}_3^{x})^2  + ({e}_3^y)^2 \Big)
 + c_2 ~\Big( { e}_3^x ~ { e}_3^y \Big) 
 + c_3~ \Big( { e}_3^x~ { e}_3^z + { e}_3^y ~ { e}_3^z  \Big).  \nn 
 \label{eq:anisotropy-1}
\eea
The magnitudes of $c_i$ ($i=1,2,3$) parameters cannot be determined on symmetry grounds, 
thus we investigate possible directions of $\hat{\bf e}_3$ for general cases of $c_i$.

%
\begin{figure}[t]
\scalebox{0.25}{\includegraphics{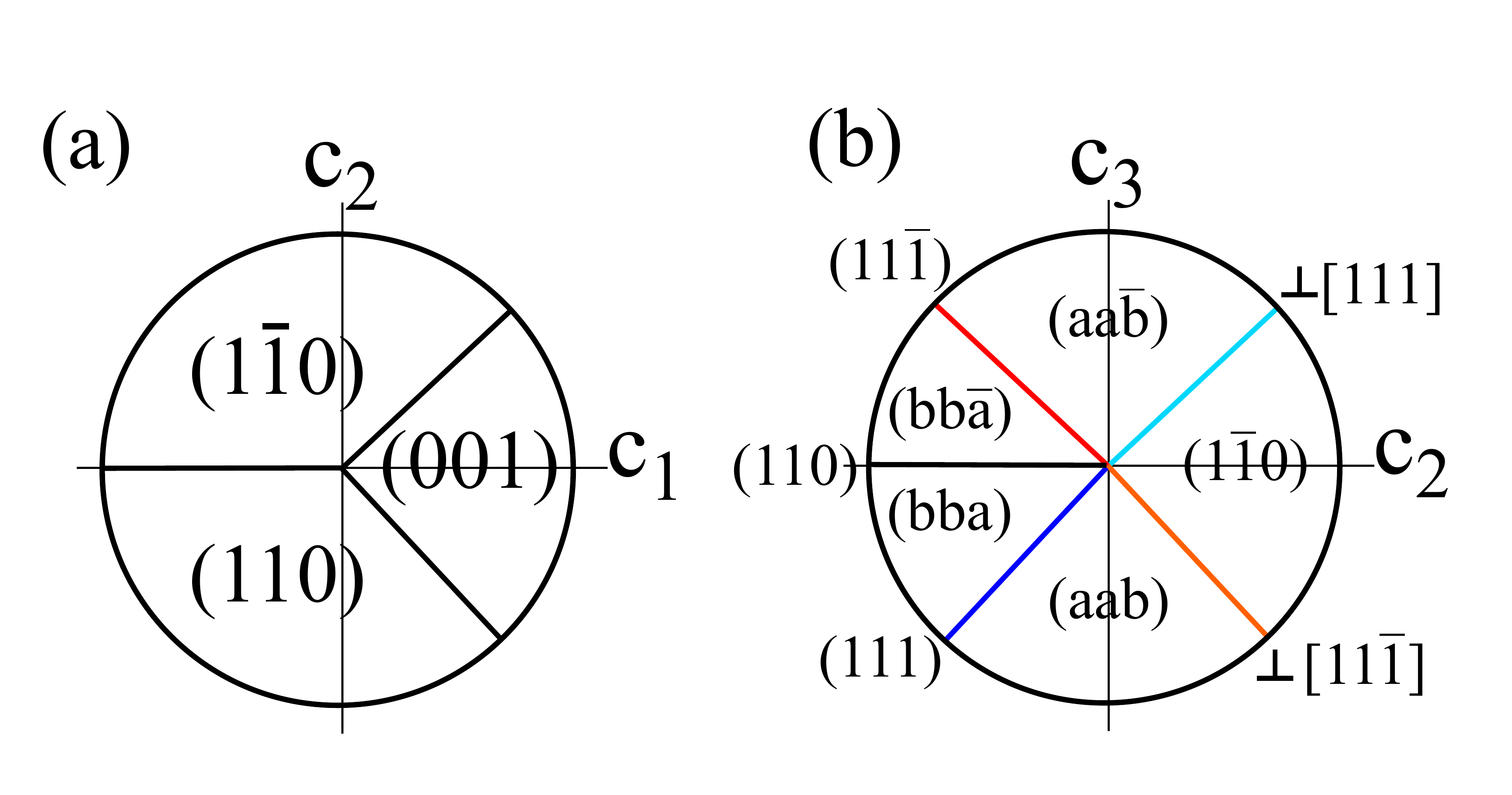}}
\caption{Directions of $\hat{{\bf e}}_3$ chosen by possible magnetic anisotropies, which are normal to the spiral plane.
(a) and (b) show the cases of the $\Gamma$Z or $\Gamma$Y spiral states and
$\Gamma$X spiral state, respectively (we take the limit $c_1=0$ for the latter case).
The symbol $\perp[111]$ indicates that any $\hat{\bf e}_3$ normal to $[111]$ is possible.
The parameters $a$ and $b$ represent continuous change of $\hat{\bf e}_3$ with $b > a$.
The $\Gamma$Z and $\Gamma$Y spiral states select (001),(110) and (1$\bar{1}$0) as spiral planes, 
whereas, for the $\Gamma$X spiral states, the spiral plane is not completely determined and there are
a number of possible choices depending on the anisotropy parameters $c_i$. }
\label{fig:aniso}
\end{figure}

For the $\Gamma$Z or $\Gamma$Y spiral states,
there are two independent parameters $c_1$ and $c_2$ that represent
possible magnetic anisotropies.
Figure~\ref{fig:aniso} (a) illustrates the chosen directions of $\hat{\bf e}_3$ that 
would minimize the free energy, depending on the relative values of $c_1$ and $c_2$.
For $c_1 >0$ and $ |c_2| <c_1$, the magnetic anisotropy chooses (001) as the 
spiral plane, {\it i.e.} $\hat{\bf e}_3 = (001)$.
For $c_2>0$ and $c_2 > c_1$, the spiral plane is (1$\bar{1}$0), 
whereas, for  $c_2 <0$ and $-c_2 > c_1$, it is (110).
Hence, one could expect that the magnetic anisotropy 
would make the spins to lie on any of (001), (110), (1$\bar{1}$0) planes
when the $\Gamma$Z or $\Gamma$Y spiral states are stabilized.

On the other hand, if the $\Gamma$X spiral state is stabilized, 
the spiral plane is not always completely determined and some ambiguity
may remain. Here one can prefer any direction as long as the components of the
unit-vector $\hat{\bf e}_3$ satisfy $|e_3^x| =| e_3^y|$,  
depending on the values of three independent parameters $c_1,~c_2$ and $c_3$.
As an example, Fig.~\ref{fig:aniso} (b) illustrates the chosen directions of $\hat{\bf e}_3$ 
for arbitrary $c_2$ and $c_3$ with $c_1 =0$. 
In this case, $\hat{\bf e}_3$ is pointing along the high symmetry directions 
only when the parameters are fine-tuned. 
For $c_3=0$, the magnetic anisotropy selects either (110) or (1$\bar{1}$0) plane
depending on the sign of $c_2$.
On the other hand, (1$\bar{1}$0) plane is chosen in an extended parameter regime of finite $c_3$ and $c_2 > c_3 $. 
For $c_2 < 0$ and $c_2 =\pm c_3$, the spiral plane becomes either (111) or (11$\bar{1}$).
For $c_2 >0 $ and $c_2 =\pm c_3$, any spiral planes can be chosen as far as their normal vector 
$\hat{\bf e}_3$ satisfies either $\hat{\bf e}_3 \cdot [111] =0 $ or $\hat{\bf e}_3 \cdot [11\bar{1}] =0$
(We use the symbols $\perp [111]$ and $\perp [11\bar{1}]$ to represent this case in Fig.~\ref{fig:aniso}). 
Beyond these special limits, the direction of the normal vector $\hat{\bf e}_3 $ deviates from 
the high symmetry lines and sensitively changes as a function of $c_i$. 
The symbols $a$ and $b$ in Fig.~\ref{fig:aniso} (b) indicate that
the direction of $\hat{\bf e}_3$ changes with $b>a$ with $a$ and $b$ being
able to change continuously.

Finally, we discuss possible future directions related to our results.
One interesting issue would be thermal order-by-disorder effects in the same Heisenberg model 
on the hyperhoneycomb lattice. Similar to the effects of quantum fluctuations, thermal fluctuations 
may also lift the classical ground state degeneracy and select special magnetic ordering patterns. 
While we show that collinear magnetic orders such as the stripy phase are promoted upon
including quantum fluctuations, it is not guaranteed that the entropic effect would also favour 
the same collinear order. In case that thermal order-by-disorder effect chooses a different state,
there might be a finite temperature transition to the entropically chosen phase at finite
temperatures due to the competition between quantum and thermal order-by-disorder effects.
Another issue is the investigation of magnetic phases in the presence of 
both anisotropic spin interactions and magnetic frustration. 
In our paper, we mainly focused on the magnetic frustration effect and studied the SU(2)-invariant 
Heisenberg model albeit we briefly discussed possible effects of magnetic anisotropies in spiral ordered
phases. On the other hand, the full understanding of the interplay between anisotropic spin interactions and 
magnetic frustration may be necessary for understanding real materials.
This would be an important topic for future theoretical studies.

\acknowledgements
We gratefully acknowledge useful discussions with Eric Kin-Ho Lee,  Subhro Bhattacharjee, and Gang Chen. 
We thank KIAS center for Advanced Computation for providing computing resources.
This work was supported by the NSERC, CIFAR, and the Supercomputing Center/Korea Institute of Science and Technology 
Information with supercomputing resources including technical support (KSC-2013-C3-035).
Some of the computations were performed on the gpc supercomputer at the SciNet HPC Consortium. 
SciNet is funded by: the Canada Foundation for Innovation under the auspices of Compute Canada; the Government of Ontario; Ontario Research Fund - Research Excellence; and the University of Toronto.
\label{sec:acknowledgements}

\appendix


\section{Hyperhoneycomb lattice structure}
\label{app-hyperhoneycomb}

\begin{figure}[t!]
\centerline{\includegraphics[width=6cm]{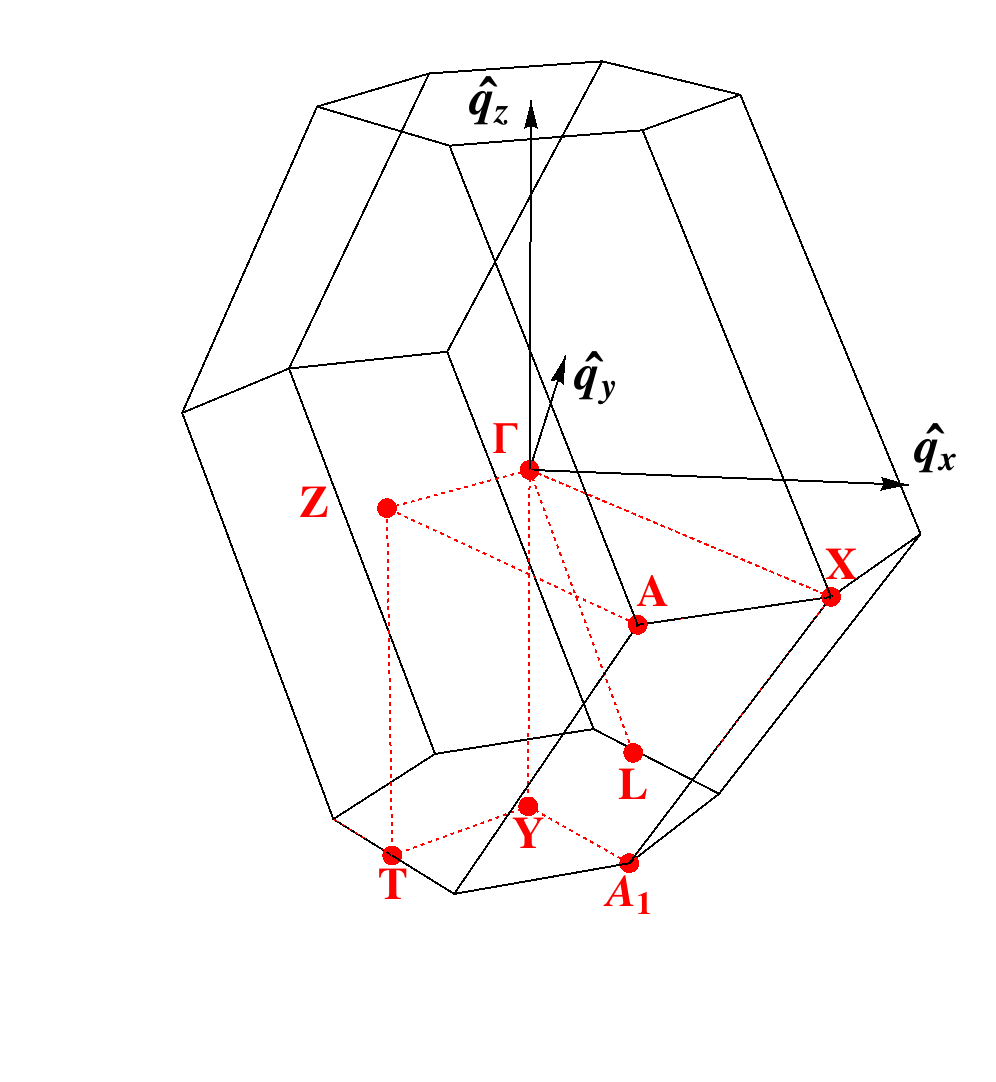}}
\caption{(Color online) Red dots indicate high symmetric points in the Brillouin zone, and their coordinates are given by $\Gamma=\left(0, 0, 0\right)$, ${\rm Z}=\left(-{\pi\over 6}, -{\pi\over 6}, 0\right)$, ${\rm T}=\left(-{\pi\over 6}, -{\pi\over 6}, -{\pi \over 2}\right)$, ${\rm X_1}=\left(-{19\pi\over 72}, -{5\pi\over 72}, -{\pi \over 2}\right)$, ${\rm Y}=\left(0, 0, -{\pi\over 2}\right)$, ${\rm A_1}=\left({11\pi \over 72}, -{11\pi \over 72}, -{\pi \over 2}\right)$, ${\rm X}=\left({29\pi \over 72}, -{29\pi \over 72}, 0\right)$, ${\rm L}=\left({\pi\over 6}, -{\pi \over 3}, -{\pi \over 4}\right)$, and ${\rm A}=\left({13\pi \over 72}, -{37\pi \over 72}, 0\right)$. Three primitive vectors of the reciprocal lattice are given by $\bs{b}_1=\left({\pi\over 3}, -{2\pi\over 3}, {\pi \over 2}\right)$, $\bs{b}_2=\left(-{2\pi \over 3}, {\pi \over 3}, -{\pi \over 2}\right)$, and $\bs{b}_3=\left({2\pi \over 3}, -{\pi \over 3}, -{\pi \over 2}\right)$, respectively.}
\label{fig:BZ-hyperhoneycomb}
\end{figure}

In this appendix, we describe the lattice structure of the three-dimensional 
hyperhoneycomb lattice.
As shown in Fig.~\ref{fig:hyperhoneycomb},
the hyperhoneycomb lattice can be regarded as a face-centered orthorhombic Bravais lattice
with a four-site basis.
The primitive lattice vectors are given by 
\bea
\bs{a}_1 = (2,4,0), ~~ \bs{a}_2 = (3,3,2), ~~ \bs{a}_3 = (-1,1,2),
\eea
and the coordinates of four basis vectors $\bs{d}_{\alpha}$ with four sublattice indices $\alpha$ ($\alpha=0,1,2,3$) are given by 
\begin{align}
\bs{d}_0\!=\!(0,0,0),~\bs{d}_1\!=\!(1,1,0), ~\bs{d}_2\!=\!(1,2,1), ~\bs{d}_3\!=\!(2,3,1).
\end{align}
The sites of four different sublattices are shown as distinct colored-spheres in Fig.~\ref{fig:hyperhoneycomb}.
The lengths of the nearest-neighbor and next-nearest-neighbor bonds are $\sqrt{2}$ and $\sqrt{6}$, respectively. 
Each sublattice site has three nearest-neighbor bonds that reside on 
a two-dimensional plane, forming $120^{\circ}$ angle to each other
as the case in the two-dimensional honeycomb lattice.
However, unlike the honeycomb lattice, three nearest-neighbor bonds in the hyperhoneycomb lattice are not 
equivalent as only two of the nearest-neighbor sites belong to the same sublattice and the remaining one does not.
One can also describe the lattice structure with the 
orthorhombic conventional unit-cell (a box in grey lines) shown in Fig.~\ref{fig:hyperhoneycomb}.
Here the lattice vectors are given by ${\bf a} = (6, 6, 0)$, 
${\bf b} = (-2,2,0)$, and ${\bf c} = (0,0,4)$ (in the same unit as used above).
The reciprocal lattice vectors are given by,
\bea
b_1 &=& \Big(  
\frac{\pi}{3} , - \frac{2\pi}{3} , \frac{\pi}{2} 
\Big) \nn \\
b_2 &=& \Big(- \frac{2 \pi}{3}, \frac{\pi}{3} , - \frac{\pi}{2}  
\Big) \nn \\
b_3 &=& \Big( 
\frac{ 2\pi}{3} , - \frac{\pi}{3} , - \frac{\pi}{2} 
\Big).
\eea
The first Brillouin zone as well as the high symmetry 
directions and points are shown in Fig.~\ref{fig:BZ-hyperhoneycomb}. 

%
%

\section{Holstein-Primakoff linear spin-wave theory}
\label{app:HP-bosons}

We discuss the computation of zero-point quantum-fluctuation energy 
within the large-$S$ semi-classical linear spin-wave theory. 
As introduced in Sec.~\ref{subsec:HP-bosons},
the Holstein-Primakoff transformation in Eq.~\eqref{eq:subsec-hp-2} 
leads to the spin wave Hamiltonian of the order ${\cal O} (S)$: 

\begin{align}
H_{\mathcal{O} ({S})} /S =& \frac{1}{2} \sum_{ij} J_{ij} \Big[ 
a_i^\dagger a_j ( \hat{\bs{x}}_i \cdot \hat{\bs{x}}_j +1 - 2 \delta_{ij} \hat{\bs{z}}_i  \cdot \hat{\bs{z}}_j )\nonumber\\
&+ \frac{1}{2} ( a_i^\dagger a_j^\dagger + a_i a_j) ( \hat{\bs{x}}_i \cdot \hat{\bs{x}}_j -1)  \Big] \nn \\
=&
\sum_{\bs{k},s}
\Big(\!
\begin{array}{cc}
\!A_{\bs{k}}^\dagger &
A_{-\bs{k}}\!
\end{array}
\!\Big)
\Big(
\begin{array}{cc}
W_{\bs{k}} & T_{\bs{k}} \\
T_{\bs{k}}^{*t} & W_{-\bs{k}}^t
\end{array}
\Big)
\Big(
\begin{array}{c}
A_{\bs{k}} \\
A_{-\bs{k}}^\dagger
\end{array}
\Big) \!+\!  J_m .
\label{eq:app-hp-1}
\end{align}
Here, the second line denotes the Fourier-transformed Hamiltonian.
$A_{\bs{k}}^\dagger = (a_{\bs{k},0}^\dagger,a_{\bs{k},1}^\dagger,a_{\bs{k},2}^\dagger,a_{\bs{k},3}^\dagger), 
~ A_{\bs{k}} = ( a_{\bs{k},0}, a_{\bs{k},1}, a_{\bs{k},2}, a_{\bs{k},3})$ represent boson creation, annihilation operators for sublattices  $0,1,2$ and $3$ at wavevector $\bs{k}$, 
and $J_m = \sum_{ij} \frac{J_{ij} }{2} \hat{\bs{z}}_i \cdot \hat{\bs{z}}_j $ is the classical energy.
$W_{\bs{k}}$ and $T_{\bs{k}} $ are $4 \times 4 $ matrices defined as
\bea
(W_{\bs{k}})^{ss'} &=& \frac{1}{16} \sum_{i \in s, j \in s'} 
\Big( J_{ij} e^{- i \bs{k} \cdot (\bs{r}_i - \bs{r}_j) } 
[ \hat{\bs{x}}_i \cdot \hat{\bs{x}}_j  \!+\!1] \!-\! 4~ \delta_{s,s'}~ J_m 
\Big), \nn \\
(T_{\bs{k}})^{ss'} &=& \frac{1}{16} \sum_{i \in s, j \in s'} \Big(  J_{ij} e^{- i \bs{k} \cdot (\bs{r}_i - \bs{r}_j) } 
[ \hat{\bs{x}}_i \cdot \hat{\bs{x}}_j  - 1 ]
\Big).
\eea
Using the Bogoliubov transformation, Eq.~\eqref{eq:app-hp-1} can be rewritten in the diagonalized basis:
\bea
H_{\mathcal{O} (S)}/S &=& \sum_{\bs{k},s} 
\Big(
\begin{array}{cc}
d_{\bs{k},s}^\dagger &
d_{-\bs{k},s}
\end{array}
\Big)
\Big(
\begin{array}{cc}
\omega_{\bs{k},s} & 0 \\
0 & \omega_{- \bs{k},s}
\end{array}
\Big)
\Big(
\begin{array}{c}
d_{\bs{k},s} \\
d_{-\bs{k},s}^\dagger
\end{array}
\Big) \nn \\
&=&  \sum_{\bs{k},s} 2 \omega_{\bs{k} ,s}~ 
\Big( d^\dagger_{\bs{k},s} d_{\bs{k},s} + \frac{1}{2} 
\Big),
\label{eq:app-hp-2}
\eea
where $\omega_{\bs{k},s} $ are the eigenvalues of 
$\Big(
\begin{array}{cc}
W_{\bs{k}} & T_{\bs{k}} \\
T_{\bs{k}}^{*t} & W_{-\bs{k}}^t
\end{array}
\Big) \cdot 
\Big(
\begin{array}{cc}
\mathbb{I}_4 & 0 \\
0 & -\mathbb{I}_4
\end{array}
\Big).
$

\section{Mean-field ans\"atze for the Schwinger boson theory\label{app:sbmft}}

Here, we describe the mean-field ans\"atze used in the Schwinger boson theory. As mentioned in Sec.~\ref{subsec:schwinger-bosons}, the ans\"atze are translationally invariant and
represented by four independent parameters $\eta_{n}~(n=1,\cdots,4)$ in a unit cell. The mean-field parameters are listed in Table \ref{tab:int-links} for six $J_1$ bonds and twelve $J_2$ bonds in a unit cell. The four parameters are defined at four different types of bonds. The parameter $\eta_{1}$ ($\eta_2$) is defined for the $J_1(z)$ ($J_1 (xy))$ bonds, where
$J_1(z)$ connects the sublattices 0 \& 1 or 2 \& 3, and $J_1 (xy)$ represents the rest of the $J_1$-type bonds.
The parameter $\eta_{3}$ is defined for the $J_2(xy-xy)$ bonds, where two sites in the bond are connected to two $J_1(xy)$ bonds.
$\eta_{4}$ is defined for the $J_2(xy-z)$ bonds, where two sites in the bond are connected to $J_1(xy)$ and $J_1(z)$ bonds. 
The above classification of the bonds is based on lattice symmetries of the hyperhoneycomb lattice. 
Under lattice symmetry operations, each type of bonds are transformed into the same type of bonds, for instance 
a $J_1(z)$ bond is transformed to another $J_1(z)$ bond. These mean-field parameters are assumed to be real numbers.

Table \ref{tab:int-links} describes two mean-field ans\"atze. Notice that the first half of $\eta_3$ bonds can have two possible signs, $\pm$. The ansatz with the positive (negative) sign describes the $\Gamma$X ($\Gamma$Y) spiral phase. 

\begin{table}[!]
\begin{ruledtabular}
\begin{tabular}{c|ccc|c}
 bond & $m$ & $n$ & ${\bf R}$ & $\eta_{ij}$ 
 \\
 \hline
 $J_1(z)$ & 0 & 1 & ${\bf 0}$ & $\eta_1$ 
 \\
 & 2 & 3 & ${\bf 0}$ & $\eta_1$
 \\
 \hline
 $J_1(xy)$ & 1 & 2 & ${\bf 0}$ & $-\eta_2$
 \\
 & 3 & 0 & ${\bf a}_1$ & $\eta_2$
 \\
 & 3 & 0 & ${\bf a}_2$ & $\eta_2$
 \\
 & 2 & 1 & ${\bf a}_3$ & $\eta_2$
 \\
 \hline
 $J_2(xy-xy)$ & 0 & 0 & ${\bf a}_1-{\bf a}_2$ & $\pm\eta_3$
 \\
 & 3 & 3 & ${\bf a}_1-{\bf a}_2$ & $\pm\eta_3$
 \\
 & 1 & 1 & ${\bf a}_3$ & $\eta_3$
 \\
 & 2 & 2 & ${\bf a}_3$ & $\eta_3$
 \\
 \hline
 $J_2(xy-z)$  & 2 & 0 & ${\bf 0}$ & $\eta_4$
 \\
 & 2 & 0 & ${\bf a}_1$ & $\eta_4$
 \\
 & 2 & 0 & ${\bf a}_2$ & $\eta_4$
 \\
 & 2 & 0 & ${\bf a}_3$ & $\eta_4$
 \\
 & 3 & 1 & ${\bf 0}$ & $\eta_4$
 \\
 & 3 & 1 & ${\bf a}_1$ & $\eta_4$
 \\
 & 3 & 1 & ${\bf a}_2$ & $\eta_4$
 \\
 & 3 & 1 & ${\bf a}_3$ & $\eta_4$
\end{tabular}
\end{ruledtabular}
\caption{Different types of bonds in a unit cell and mean-field ans\"atze $\{\eta_{ij}\}$ for the Schwinger boson mean-field theory. In the table, $i=({\bf 0} ; m)$ and $j=({\bf R} ; n)$, where the first component means the lattice vector and the second the sublattice. The ansatz with the positive (negative) sign for the first half of $\eta_3$ bonds describes the $\Gamma$X ($\Gamma$Y) spiral phase. 
\label{tab:int-links}}
\end{table}

\section{Symmetry analysis of magnetic anisotropy}
\label{app-symmetry}

Here, we briefly discuss the symmetries of the hyperhoneycomb lattice following Ref.\onlinecite{PhysRevB.89.045117},
and derive the symmetry transformation properties of the order parameter ${\bf d}$, 
defined in Eq.~\eqref{eq:sQ-1}.
The ideal hyperhoneycomb lattice is described by the space group $Fddd$ and the point group D$_{2h}$. 
There are three types of symmetry operations in the hyperhoneycomb lattice 
(See Fig.~\ref{fig:hyperhoneycomb}):
\begin{enumerate}
\item Inversion at the bond center of sublattices 1 and 2, or sublattices 0 and 3. 
\item Three orthogonal $C_2$ axes at the bond center of sublattices 0 and 1, 
or sublattices 2 and 3. These axes are parallel to the lattice vectors  ${\bf a},~{\bf b}$ 
and ${\bf c}$ of the underlying face-centered orthorhombic Bravais lattice 
(defined in Appendix~\ref{app-hyperhoneycomb}). 
\item Glide planes parallel to each face of the orthorhombic unit-cell followed by translation ${\bf a}_i/2$ 
(defined in Appendix~\ref{app-hyperhoneycomb}).
\end{enumerate}

\noindent There are four generators for D$_{2h}$ point group: inversion $\mathcal{I}$ and 
three $C_2$ rotations, $C_{2{\bf a}} $, $C_{2 {\bf b}}$, $C_{2 {\bf c}}$. 
Using these four generators, one can examine how the spin ${\bf S}_i$ [Eq.~\eqref{eq:sQ-1}], 
that is a pseudo-vector, transforms under each symmetry operation. 

We consider two different types of spiral states:
$\Gamma$Z spiral state with the ordering wavevector ${\bf Q} = q(1,1,0)$ 
and $\Gamma$XY spiral state with the ordering wavevector, either
${\bf Q} = q( 1, {\bar{1}}, 0)$ or ${\bf Q} = q(0,0,1)$.
For $0.17< J_2/J_1 <0.5$, the ground sate is the $\Gamma$Z spiral state 
with the ordering wavevector ${\bf Q} = q(1,1,0)$ 
and the corresponding little group (for the remaining lattice symmetries) is generated by 
$\mathcal{I},~ C_{2 {\bf a}}$. 
Under the symmetry operations, the order parameter ${\bf d}$ 
transforms as
\bea
\mathcal{I} &:& {\bf d} \rightarrow {\bf d}^* \nn \\
C_{2 {\bf a}} &:& d_x \rightarrow d_y,~ d_y \rightarrow d_x, ~ d_z \rightarrow - d_z. 
\eea 
Similarly, for $J_2/J_1 > 0.5$, the ground state is the $\Gamma$XY spiral state
with the ordering wavevector, either 
${\bf Q} = q( 1, {\bar{1}}, 0)$ or ${\bf Q} = q(0,0,1)$.
For the case of the ordering wave vector ${\bf Q} = q( 1, {\bar{1}}, 0)$, 
the little group is generated by $\mathcal{I}, ~C_{2 {\bf b}}$
and the order parameter {\bf d} transforms as
\bea
\mathcal{I} &:& {\bf d} \rightarrow {\bf d}^* \nn \\
C_{2 {\bf a}} &:& d_x \rightarrow -d_y,~ d_y \rightarrow -d_x, ~ d_z \rightarrow - d_z. 
\eea
In the case of the ordering wavevector ${\bf Q} = q(0,0,1)$, 
the little group is generated by $\mathcal{I}, ~ C_{2 {\bf c}}$. 
The order parameter {\bf d} transforms under these symmetries as follows.
\bea
\mathcal{I} &:& {\bf d} \rightarrow {\bf d}^* \nn \\
C_{2 {\bf c}} &:& d_x \rightarrow -d_y,~ d_y \rightarrow -d_x, ~ d_z \rightarrow d_z. 
\eea
Based on these symmetry-transformation properties of the order parameter {\bf d}, 
one can figure out the symmetry-allowed Landau free energy 
in terms of ${\bf d}$ as shown in Eq.~\eqref{eq:anisotropy-1}.

\bibliography{J1-J2-hyperhoneycomb}

\end{document}